\documentclass[aoas]{imsart}

\RequirePackage{amsthm,amsmath,amsfonts,amssymb}
\RequirePackage[authoryear]{natbib}
\RequirePackage[colorlinks,citecolor=blue,urlcolor=blue]{hyperref}
\RequirePackage{graphicx}
\RequirePackage[table,dvipsnames]{xcolor}
\RequirePackage{booktabs}
\RequirePackage{floatpag}
\RequirePackage{algorithm}
\RequirePackage{algorithmicx}
\RequirePackage{algpseudocode}
\usepackage[format=plain,
font=it]{caption}
\DeclareCaptionLabelFormat{andtable}{#1~#2  and  \tablename~\thetable}

\let\oldtabular\tabular
\let\endoldtabular\endtabular
\renewenvironment{tabular}{\rowcolors{4}{white}{trevorblue!15}\oldtabular}{\endoldtabular}

\startlocaldefs
\definecolor{lightgrey}{rgb}{0.9,0.9,0.9}
\definecolor{darkgreen}{rgb}{0,0.3,0}

\definecolorset{rgb}{}{}{darkred,0.8,0,0;darkgreen,0,0.5,0;darkblue,0,0,0.5}

\newcommand{\dx}{\mbox{d}}

\renewcommand{\vec}[1]{\mathbf{#1}}
\newcommand{\numTaxa}{N}
\newcommand{\numTraits}{D}

\newcommand{\traitData}{\vec{Y}}

\newcommand{\latentData}{\vec{X}}
\newcommand{\latentdata}{\vec{x}}
\newcommand{\latentDatum}{x}

\newcommand{\distanceMatrix}{\mathbf{Y}}
\newcommand{\distance}{y}
\newcommand{\summant}{r}

\newcommand{\cdensity}[2]{\ensuremath{p(#1 \,|\,#2)}}
\newcommand{\density}[1]{\ensuremath{p(#1 )}}

\newcommand{\mdsSD}{\sigma}
\newcommand{\mdsVariance}{\mdsSD^2}

\newcommand{\modelDistance}{\delta}

\newcommand{\normalCDF}[1]{\Phi \left( #1 \right)}

\definecolor{lightgrey}{rgb}{0.9,0.9,0.9}
\definecolor{darkgreen}{rgb}{0,0.3,0}

\definecolorset{rgb}{}{}{darkred,0.8,0,0;darkgreen,0,0.5,0;darkblue,0,0,0.5}

\definecolor{trevorblue}{rgb}{0.330, 0.484, 0.828}
\definecolor{trevoryellow}{rgb}{0.829, 0.680, 0.306}

\endlocaldefs

\begin{document}

\begin{frontmatter}
\title{Bayesian mitigation of spatial coarsening for a Hawkes model applied to gunfire, wildfire and viral contagion}
\runtitle{Bayesian mitigation of spatial coarsening for a Hawkes model}

\begin{aug}
\author[A]{\fnms{Andrew J.} \snm{Holbrook}\ead[label=e1]{aholbroo@g.ucla.edu}},
\author[B]{\fnms{Xiang} \snm{Ji}\ead[label=e2]{xji4@tulane.edu}}
\and
\author[C]{\fnms{Marc A.} \snm{Suchard}\ead[label=e3]{msuchard@ucla.edu}}
\address[A]{UCLA Biostatistics, \printead{e1}}

\address[B]{Department of Mathematics, Tulane University, \printead{e2}}

\address[C]{UCLA Biostatistics, Human Genetics and Computational Medicine, \printead{e3}}
\end{aug}

\begin{abstract}
Self-exciting spatiotemporal Hawkes processes have found increasing use in the study of large-scale public health threats ranging from gun violence and earthquakes to wildfires and viral contagion.  Whereas many such applications feature locational uncertainty, i.e., the exact spatial positions of individual events are unknown, most Hawkes model analyses to date have ignored spatial coarsening present in the data.
Three particular 21st century public health crises---urban gun violence, rural wildfires and global viral spread---present qualitatively and quantitatively varying uncertainty regimes that exhibit (a) different collective magnitudes of spatial coarsening, (b) uniform and mixed magnitude coarsening, (c) differently shaped uncertainty regions and---less orthodox---(d) locational data distributed within the `wrong' effective space.
We explicitly model such uncertainties in a Bayesian manner and jointly infer unknown locations together with all parameters of a reasonably flexible Hawkes model, obtaining results that are practically and statistically distinct from those obtained while ignoring spatial coarsening. This work also features two different secondary contributions: first, to facilitate Bayesian inference of locations and background rate parameters, we make a subtle yet crucial change to an established kernel-based rate model; and second, to facilitate the same Bayesian inference at scale, we develop a massively parallel implementation of the model's log-likelihood gradient with respect to locations and thus avoid its quadratic  computational cost in the context of Hamiltonian Monte Carlo.  Our examples involve thousands of observations and allow us to demonstrate practicality at moderate scales.
\end{abstract}


\end{frontmatter}


\section{Introduction}\label{sec:intro}

Spatiotemporal Hawkes processes  \citep{reinhart2018review} are stochastic point processes that have found use in the modeling of various \emph{self-excitatory} phenomena in space and time such as earthquakes and their aftershocks \citep{hawkes1973cluster,ogata1988statistical,zhuang2004analyzing,fox2016spatially}, retaliatory gun violence \citep{loeffler2018gun,park2019investigating,holbrook2021scalable}, wildfires \citep{schoenberg2004testing} and viral epidemics \citep{kim2011spatio,meyer2014power,choi2015constructing,rizoiu2018sir,kelly2019real}.  These applications all share at least one characteristic that, after observing an event, one expects to observe one or more events nearby and soon after.
Because spatial proximity to an event increases the probability of observing another event, accurate model inference hinges on precise locational data.

Unfortunately, noisy, incomplete or otherwise coarsened spatial data seem to be the norm in many Hawkes process applications.  Urban gunfire data sources may provide location data at city block precision or rounded to the nearest 100 meters \citep{holbrook2021scalable}.  Scientists estimate the spatial position of an earthquake from noisy seismic wave energy arrival times at remote stations \citep{lomax2009earthquake}.  Thus, the recording of seismic locations amounts to an inverse problem arising from the use of remote sensing and complex physical models.  In the best scenario and due to privacy concerns, viral case registries provide the hospital or medical clinic that receives the sickened patient, an imprecise stand-in for the location at which the patient first contracts the virus.  More often, epidemiological data arise from heterogeneous public health sources that make use of varying levels of spatial precision, be they on the national, provincial or municipal level \citep{park2018non,holbrook2021massive}.
We account for such spatial coarsening by directly incorporating locational uncertainty into our model in the form of prior distributions on spatial positions of individual events.  The upshot is a Bayesian hierarchical model with global structure and event-specific prior distributions dictated by the weaknesses of the data at hand.  Section \ref{sec:priors} discusses these priors and how they relate to the general theoretical framework for coarsening established in \citet{heitjan1991ignorability}.

We demonstrate our approach with three distinct 21st century public health crises, each featuring its own particular spatial uncertainty and scope.  We first consider Washington D.C.~gunfire data generated throughout the span of 2018.  Here, the Government of the District of Columbia has purposefully rounded each gunshot's latitudinal and longitudinal coordinates to an effective 100 meter precision.  Because this data originate from a spatially precise acoustic gunshot location system (AGLS) \citep{loeffler2018gun}, a reasonable prior on the spatial position of each gunshot is a uniform distribution on the 10,000 $m^2$ square centered at the observed data.  Alaskan wildfire data from the years 2015 to 2019 feature a different kind of spatial uncertainty.  Each observation features approximate spatial coordinates of the fire at the time of discovery as well as the fire's size, in acres, at the time of discover.  Because we do not know the direction of each fire's expansion at the time of discovery, the principle of indifference suggests that we model each wildfire's ignition location as taking position with equal probability within a circle centered at the given discovery coordinates but with area matching the discovery's acreage.  In contrast to the Washington D.C. gunfire example, this application provides for differential spatial uncertainty between events.

A third application, the global spread of influenza from 2000 to 2012, presents a radically different flavor of spatial bias.
Because spatial proximity to an event increases the probability of observing another event, the statistician that employs a spatiotemporal Hawkes process must take care to adequately define spatial relationships between locations in a way that takes the nature of the target phenomenon into account.
Viruses spread across immensely complex human networks shaped by our relationships, institutions and economies.  On the global scale, human air transportation networks capture the majority of viral transmission between geographic locations \citep{brockmann2013hidden}.  Thus, to model global viral spread, one must build information about these transportation networks into one's model.  Failing to do so may lead to biased results that deliver incorrect insights into a crucial global public health challenge. This difficulty could be one of the reasons that the spatiotemporal Hawkes process has not found use for modeling global viral transmission.
Another reason for such a hole in the literature is that global epidemiological data often arise from heterogeneous public health sources that make use of varying levels of locational precision.  Since spatial nearness is a primary datum for the spatiotemporal Hawkes process, it is essential that our conception of nearness be coherent. How far is Beijing from China? How far is California from France?  We would like to avoid such questions as well as mixed-methodological approaches such as randomizing locations labels to, say, cities according to some contrived weighting scheme prior to analysis.

We must, therefore, use an expressive prior to simultaneously account for these two sources of spatial bias. Bayesian multidimensional scaling \citep{desarbo1998bayesian,oh2001bayesian,oh2007model,holbrook2021massive} probabilistically maps from pairwise global air transportation distances between countries to random variables within a latent Euclidean space, while our spatiotemporal Hawkes model describes the spread of viral cases, the locations of which are the very same low-dimensional latent variables.  For viral case data arising from the same country, the temporal information provided by the Hawkes process efficiently informs the distribution of latent locations on a finer, domestic scale.

In meeting our goal of joint and fully Bayesian inference over location variables and model parameters, we must develop a model for the Hawkes process background rate that admits posterior inference for its individual parameters while retaining flexibility.  As a secondary contribution, we develop just such a novel background rate model and use MCMC to compute posterior distributions for all spatiotemporal Hawkes process parameters, a first in the presence of a non-trivial background rate.   Another secondary contribution, we have made significant additions to the \textsc{hpHawkes} \textsc{R} package \url{https://github.com/suchard-group/hawkes} to facilitate high performance computing for posterior distributions of Hawkes process locations.  In particular, we have developed a fully parallelized implementation of the Hawkes log-likelihood gradient with respect to spatial locations (Appendix \ref{sec:parallelization}).

\section{Modeling}

\newcommand{\x}{\mathbf{x}}
\newcommand{\dd}{\mbox{d}}
\newcommand{\Id}{\mathbf{I}}
\newcommand{\one}{\boldsymbol{1}}

The strategy we use to model our three different target applications is to specify a single, adequately flexible data generative process in the form of a spatiotemporal Hawkes model (Section \ref{sec:hawkesModel}) and to design priors on event locations based on spatial biases encoded in each application's data (Section \ref{sec:priors}).

\subsection{Spatiotemporal self-excitation}\label{sec:hawkesModel}
\newcommand{\ttimes}{\mathbf{t}}

The spatiotemporal Hawkes process is an inhomogeneous Poisson point process \citep{daley2003introduction,daley2007introduction} model for random variables $(\x,t)\in \mathbb{R}^D\times \mathbb{R}^+$ in space and time, where the intensity function
\begin{align*}
\lambda(\x,t) = \mu(\x,t) + \xi(\x,t) = \mu(\x,t) + \sum_{t_n<t} g(\x - \x_n, t - t_n)
\end{align*}
describes the infinitesimal rate conditioned on all other observations $(\x_n,t_n)$ for $n=1,\dots,N$ and  $\x_n = (\latentDatum_{n1}, \ldots, \latentDatum_{n \numTraits} )$.
Here, $\mu(\cdot,\cdot)$ is the background rate symmetric in time, $\xi(\x,t)$ the self-excitatory rate and $g(\cdot,\cdot)$ a triggering function determining the self-excitatory behavior of the process. As in  \cite{mohler2014marked,loeffler2018gun,holbrook2021scalable}, we specify a triggering function that is exponential in time and Gaussian in space
\begin{align*}
\xi(\x,t) = \frac{\theta \omega}{ h^D} \sum_{t_n<t} e^{- \omega\, (t-t_n) }  \phi\left(\frac{\x-\x_n}{h}\right) \, ,
\end{align*}
where $\omega$, $h$ and $\theta$ are strictly positive parameters.  Similar to \cite{holbrook2021scalable} but with the inclusion of the indicator function $\mathcal{I}_{[t\neq t_n]}$ a key difference, we use a flexible Gaussian kernel smoother for the endemic rate
\begin{align*}
\mu(\x,t) = \frac{\mu_0}{\tau_x^D\,\tau_t} \sum_{n=1}^N \phi\left(\frac{\x-\x_n}{\tau_x}\right) \, \phi\left(\frac{t-t_n}{\tau_t}\right)  \mathcal{I}_{[t\neq t_n]} \, ,
\end{align*}
where the indicator function efficiently ensures that events do not contribute to their own probability of occurrence (see joint probability density function Equation \ref{eq:likelihood}, below), representing a novel and necessary departure from \cite{holbrook2021scalable} if one wishes to infer background rate parameters and process locations in a Bayesian fashion. We call $1/\omega$ and $h$ and $\tau_t$ and $\tau_x$ the self-excitatory and background lengthscales or bandwidths, respectively. We call $\mu_0$ and $\theta$ the background and self-excitatory rate weights, and their relative magnitudes determine the amount of self-excitatory behavior exhibited by the process.  With $\Theta = (\mu_0,\tau_x,\tau_t,\theta,\omega,h)$, the likelihood \citep{daley2003introduction} for data $(\latentData,\ttimes)=\left((\x_1,t_1), ..., (\x_N,t_N)\right)$ is
\begin{align*}
\mathcal{L}(\latentData,\ttimes|\Theta)
= \exp \left( - \int_{\mathbb{R}^D} \int_0^{t_N} \lambda(\x,t) \, \dd t\, \dd\x  \right)  \prod_{n=1}^N \lambda(\x_n,t_n) := e ^{ - \Lambda(t_N) } \cdot \prod_{n=1}^N \lambda_n  \, .
\end{align*}
Although integrating over the entirety of $\mathbb{R}^D$ rather than a relevant subset is a popular and often necessary modeling decision, one must regard this choice as an approximation when measurement over $\mathbb{R}^D$ is incomplete \citep{schoenberg2013facilitated}.  This fact will provide an additional argument for our proposed modeling approach in Section \ref{sec:bmdsHawkes}.
The background rate's indicator function does not change the integration term, so $\Lambda(t_N)$ is the same as in \cite{holbrook2021scalable}:
\begin{gather*}
\Lambda(t_N) = \mu_0 \sum_{n=1}^N \left(\Phi\left(\frac{t_N-t_n}{\tau_t} \right) -\Phi\left(\frac{-t_n}{\tau_t} \right) \right) - \theta \sum_{n=1}^N \left( e^{-\omega\, (t_N-t_n)} -1 \right)  \\ \nonumber
= \sum_{n=1}^N \left(  \mu_0\left(  \Phi\left(\frac{t_N-t_n}{\tau_t} \right) -\Phi\left(\frac{-t_n}{\tau_t} \right)\right)- \theta \left( e^{-\omega\, (t_N-t_n)} -1 \right) \right)  := \sum_{n=1}^N \Lambda_n \, .
\end{gather*}
Taken together, the log-likelihood is
\begin{align}\label{eq:likelihood}
\ell(\latentData,\ttimes|\Theta) &= - \Lambda(t_N) + \sum_{n=1}^N\log \lambda_n   \\ \nonumber
&= \sum_{n=1}^N\Bigg\{ \log \Bigg[  \sum_{n'=1}^N \Bigg( \frac{\mu_0\, \mathcal{I}_{[t_n\neq t_{n'}]}}{\tau_x^D\,\tau_t} \phi\left(\frac{\x_n-\x_{n'}}{\tau_x}\right)  \phi\left(\frac{t_n-t_{n'}}{\tau_t}\right) \\ \nonumber
& \hspace{9em} +\frac{\theta \omega\, \mathcal{I}_{[t_{n'}<t_n]}}{ h^D}e^{- \omega\, (t_n-t_{n'})}   \phi\left(\frac{\x_n-\x_{n'}}{h}\right) \Bigg)\Bigg]   - \Lambda_n \Bigg\} \\ \nonumber
&:= \sum_{n=1}^N \left[
\log \left(  \sum_{n'=1}^N \lambda_{nn'} \right)  - \Lambda_n \right] := \sum_{n=1}^N \ell_n  \, .
\end{align}
In all three applications, we equip $\mu_0$ and $\theta$ with standard normal priors truncated to be greater than 0..  In contrast to \citet{mohler2014marked,loeffler2018gun,holbrook2021scalable}, we perform joint inference on all model lengthscales.  To do so, we lend truncated normal priors to all model inverse lengthscales.  We maintain constraints $0<1/\omega<\tau_t$ and $0<h<\tau_x$.  Finally, we set the prior standard deviations of the background inverse lengthscales to be 10-times those of their respective self-excitatory counterparts.  In this way, we encode our general expectation that self-excitation occurs at a finer scale than that of the background process.  Regardless, we find that the thousands of observations present in each of our applications can easily overpower the soft prior constraints given by the prior standard deviations (Table \ref{tab::wildfires}).

Importantly, our model is similar to that of \citet{mohler2014marked}, who finds that the spatial lengthscales $h$ and $\tau_x$ may sometimes be exchanged with only a small change to the likelihood.  For this reason, \citet{mohler2014marked} fixes the two parameters to be equal to avoid multimodality.  Whereas our priors help ameliorate this issue, they do not solve it.  Amazingly, we find that inferring locations as discussed in the following section can actually help solve this problem (Table \ref{tab::wildfires}).  As an upshot, we are able to retain full model flexibility.  This is all the more important because \citet{reinhart2018self} show that poorly estimated background processes contribute to biased estimates of self-excitation.

\subsection{Modeling spatial uncertainty}\label{sec:priors}

\begin{figure}[t!]
	\centering
	\includegraphics[width=0.6\linewidth]{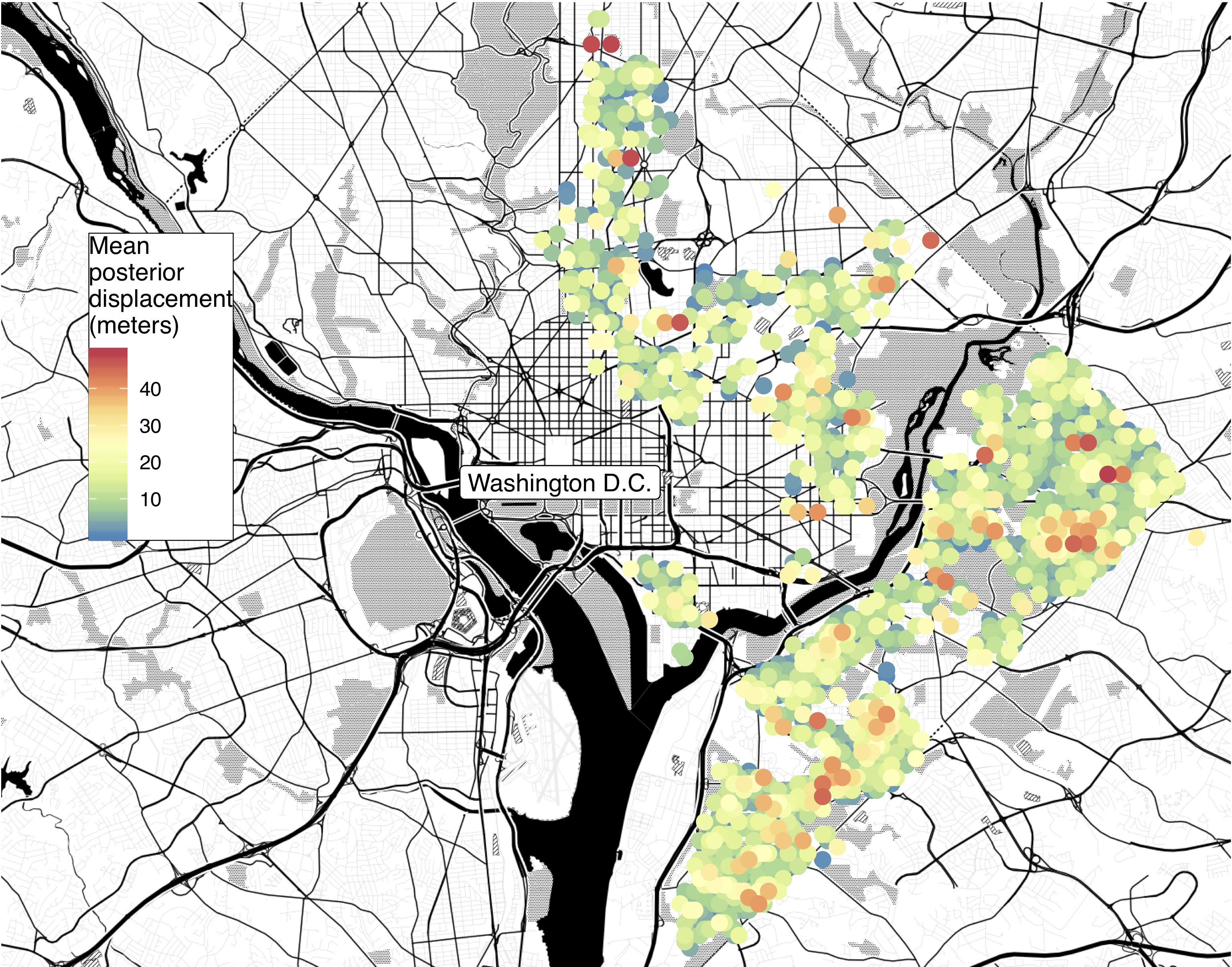}
	\caption{We color the observed locations $\mathfrak{x}_n$ of 3,982 gunshots occurring in the year 2018 by the magnitude of the mean posterior displacement of each event's inferred locations $\x_n$: $||\sum_{s=1}^S (\x_n^{(s)}-\mathfrak{x}_n )/S||_2$, where $S$ is the total number of MCMC states.  For each event, this measure communicates the amount of posterior displacement in a single, general direction away from the observed location $\mathfrak{x}_n$.}\label{fig:dcmap}
\end{figure}

Whereas we apply the same likelihood to each of our target applications, we must craft our priors on individual event locations in a way that respects the phenomenon being modeled and the spatial coarsening that gives rise to each specific dataset. Rather than obtaining observations that belong to the sample space of our random variable of interest, we observe \emph{coarse} data \citep{heitjan1991ignorability,heitjan1993ignorability} within the power set of that sample space.  Rounded, heaped, truncated, censored and missing data are just a few common examples of coarsening.  If one knows the coarsening mechanism and it is not stochastic, then the data are \emph{grouped}.  Rounding and truncation with fixed precision are prominent examples of grouping.  Addressing rounding is as simple as integrating the originating likelihood over the uncertainty region the rounding induces. Adopting the established parlance of missing data, \citet{heitjan1991ignorability} show that a stochastic coarsening mechanism is \emph{ignorable} if (a) it is coarsened at random (CAR) and (b) the parameters of the data generating and coarsening processes are distinct.  If one assumes that ignorability holds, then modeling the data as grouped, i.e., ignoring the stochasticity of the coarsening mechanism, is completely valid. In this paper, we use the phrases spatial coarsening, spatial uncertainty and even spatial bias interchangeably.


In the remainder, $\x_n$ continues to denote an individual location that interfaces directly with the Hawkes model likelihood. This is a location variable.  We denote its corresponding observed locational datum as $\mathfrak{x}_n$ and let $\mathfrak{X}$ be the collection of all $N$ observed locations.

\subsubsection{Washington D.C.~gun violence}\label{sec:dc_uncert}

We first apply our Hawkes model to analyze gunfire in the American capital throughout the year 2018.  The data feature 3,982 gunshots obtained from a spatially precise AGLS (Section \ref{sec:intro}) but with latitudinal and longitudinal coordinates rounded to the nearest three decimal points for the purpose of privacy.  This rounding amounts to recording observations within an approximate 100 meter precision in localized vertical and horizontal axes.  Due to the precision of the original AGLS data, we are confident in specifying uniform priors over the 10,000 $m^2$ square centered at each location for each location, i.e., in local coordinates scaled to meters:
\begin{align}
p(\x_n)  \propto 1 \, , \quad  \mathfrak{x}_{nd} - 50  < \x_{nd} < \mathfrak{x}_{nd} + 50\, ,  \quad n=1,\dots, N, \: d=1,2 \, . \label{eq:locsPrior1}
\end{align}
Our uncertainty is uniform in both shape and magnitude throughout the sample.  As stated above, rounding is an example of grouping, and we know that our prior specification is valid and corresponds directly to inference based on the \emph{grouped-data likelihood} of \citet[Example 1]{heitjan1991ignorability}. In other words, our latent variable formulation accounts for grouping by integrating over the region of uncertainty induced by the grouping mechanism.   Failing to account for this grouping leads to biased inference.

\subsubsection{Alaskan wildfire ignitions}

Next, we model the occurrence and spread of 2,925 wildfires in Alaska through the years 2015 to 2019.  Specifically, we would like to use the exact time and place of ignition for each wildfire as our data.  Instead, we have the time, rough spatial coordinates $\mathfrak{x}_n$ and area $A_n$ of the fire at discovery.  Because we do not know the direction of each wildfire's extent, we invoke the principle of indifference \citep{marquis1825essai} and assume equal extent in all directions, i.e., that each uncertainty region is a circle centered at the given coordinates, assuming effects of geography are negligible.  Here, we specify the radius $r_n$ of each circular uncertainty region so that the circle's area matches the size of the wildfire at time of discovery:
\begin{align}
p(\x_n) \propto 1 \, , \quad  ||\x_n - \mathfrak{x}_n||_2 < r_n = \sqrt{A_n/\pi}\, , \quad n=1,\dots,N,\: D=2\, . \label{eq:locsPrior2}
\end{align}
This example,therefore, stands in contrast to the gun violence example insofar as the shape of uncertainty regions are circular rather than square, and the magnitudes of these circles vary across all observations.  The coarsening mechanism appears to be random, but it is impossible to capture the complicated processes that lead a passerby to discover a wildfire at any particular extent.  Unlike the D.C.~gunfire example, we do not know the exact circumstances that bring about the data's observed spatial coordinates, so we must simply assume that the CAR condition holds. Less problematic is the assumption that data generating and data coarsening mechanisms have distinct parameters.  Having arrived at ignorability, our prior specification again corresponds to valid inference based on the grouped-data likelihood.

\subsubsection{Global influenza contagion}\label{sec:bmdsHawkes}

\begin{figure}[t!]
	\centering
	\includegraphics[width=0.7\linewidth]{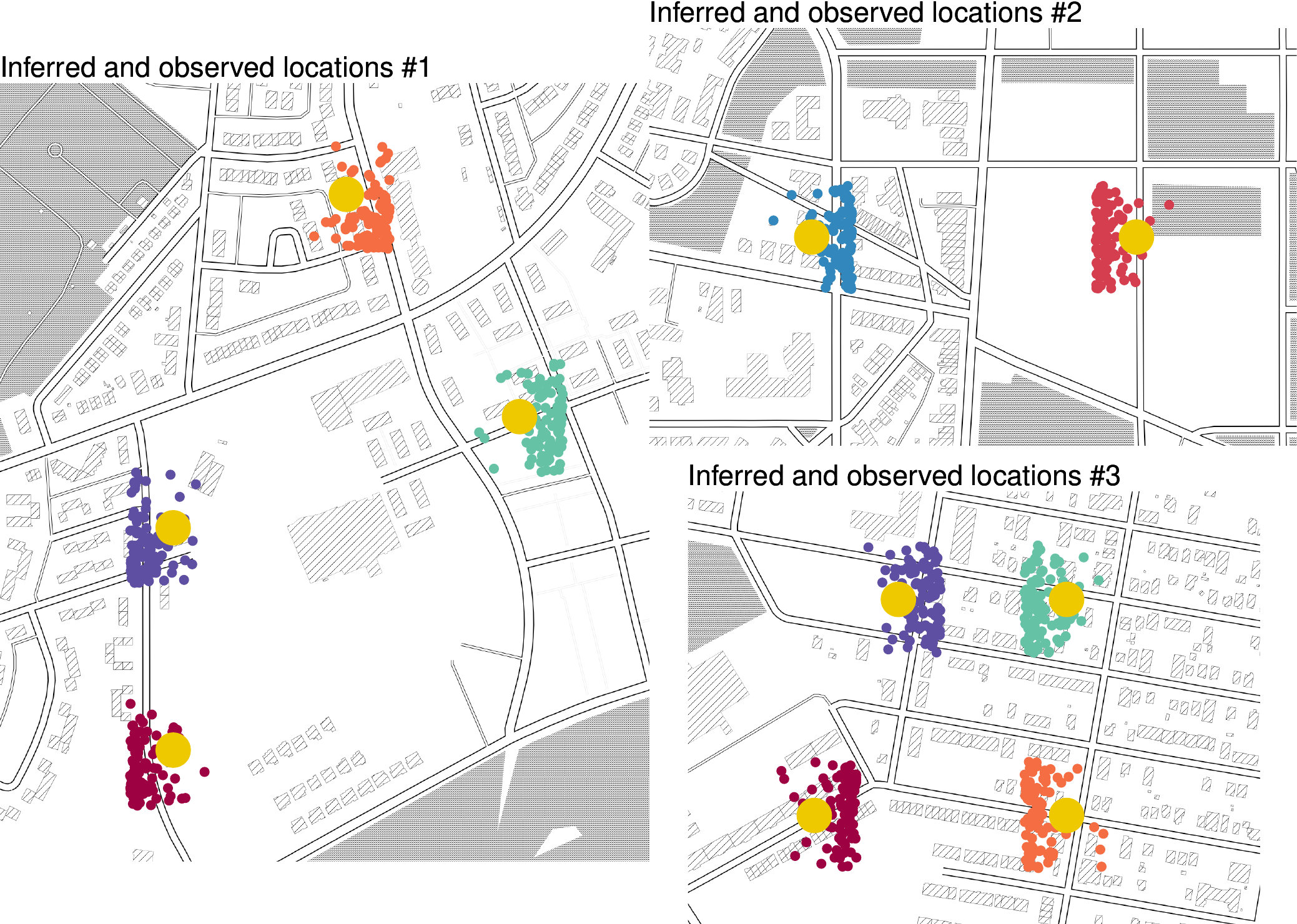}
	\caption{Visualizing the relationship between observed locations $\mathfrak{x}_n$ (yellow) and posterior sample locations $\x_n^{(s)}$ for ten gunshots in the District of Columbia.  Inferred locations may deviate from observed for multiple reasons. In the first plot, differences in date and time range from 11 to 120 hours; in the second, the two events differ by 55 days. On the one hand, the gunshots in the first plot occur in a gunfire dense area but separated by a low-activity shopping center.  On the other hand, the gunshots in the second plot are spatially isolated from other events.}\label{fig:dcmapZoom}
\end{figure}

\paragraph{Doubly debiased inference}

4,733 influenza cases collected from 64 countries worldwide between 2000 and 2012 provide a much more difficult modeling task.  Approximately 1/3 of the observations bear labels for the city; 1/3, the province or state; and 1/3, the country in which the case occurred.  We would like to proceed in a similar manner as with the gun violence and wildfires analyses, but restricting location variables to the complicated borders of countries, provinces or cities is technically infeasible.  Furthermore, naive spatial distances between locations on planet Earth fail to capture the way viruses propagate around the globe.  The global human air transportation network, specifically the number of humans traveling between locations, provides a much better tool for tracking the spread of viral strains \citep{brockmann2013hidden}.  We therefore propose to model the locations of each viral case in such a way that simultaneously accounts for the multi-precision nature of the data and the outsized role played by human air transport.

Classical multidimensional scaling (MDS) is a two-step method for mapping  from pairwise distances or dissimilarities between objects to representations of these objects within a low-dimensional Euclidean space \citep{kruskal1964multidimensional}. In modeling the global spread of influenza, we let $\traitData=\traitData(\mathfrak{X})$ be an $N\times N$ matrix of pairwise distances $\distance_{nn'}$ generated by \citet{brockmann2013hidden} and inversely related to the number of air traffic passengers exchanged between the countries where cases $n$ and $n'$ occurred. Given any such matrix $\traitData$, the centering transformation
\begin{align*}
\traitData \longmapsto - \frac{1}{2} \left(\Id-\frac{1}{N}\one\one^T\right)  \traitData^{\circ 2}  \left(\Id-\frac{1}{N}\one\one^T\right)
\end{align*}
results in a positive semi-definite matrix corresponding to the sample covariance of $N$ points existing in some $D$-dimensional subspace of $N$-dimensional Euclidean space.  After obtaining this sample covariance, a simple application of principal component analysis (PCA) \citep{pearson1901liii} renders a low-dimensional representation of the $N$ objects of interest.
On the one hand, objects with smaller pairwise distances arrange themselves closer in $L_2$ distance within the low-dimensional space than objects with larger pairwise distances, leading to interpretable visualizations.  On the other hand, (1) both the centering transformation and the eigendecomposition of PCA scale $O(N^3)$ in computational complexity, (2) low-dimensional representations fail to communicate uncertainty arising from randomness in the data generating mechanism and (3) secondary modeling of the low-dimensional representations results in difficult to quantify dependencies on the MDS process.

Bayesian MDS (BMDS) offers a way around these problems by positing that each object's latent location is a random variable, translating the MDS projection into a probability model on the observed pairwise distances conditioned on distances between latent locations \citep{ramsay1982some} and specifying an appropriate prior distribution over these locations.
For any two distinct objects $n$ and $n'$, we follow \citet{oh2001bayesian} and model their observed pairwise distance as conditionally independent, truncated normal random variables
\begin{align*}
\distance_{nn'} \sim \mbox{N} \left( \modelDistance_{nn'}, \mdsVariance \right) \mathcal{I}_{[\distance_{nn'}  > 0]}  \text{ for } n > n',
\end{align*}
where the centrality parameter $\modelDistance_{nn'} = || \latentdata_n - \latentdata_{n'} ||_2$ is the Euclidean distance between latent locations $\latentdata_n$  and $\latentdata_{n'}$
in $\mathbb{R}^{\numTraits}$.
Conditioned on all latent locations $\latentData$, the probability density function of observed distance data $\distanceMatrix$ becomes
\begin{align}
\cdensity{\traitData}{\latentData, \mdsVariance} & \propto
\left(
\mdsVariance
\right)
^{\frac{\numTaxa ( 1-\numTaxa) }{4}}
\exp
\left(	-
\sum_{n > n'}
\summant_{nn'}
\right)
\nonumber \\
\summant_{nn'} &=
\frac{ \left( \distance_{nn'} - \modelDistance_{nn'} \right)^2 }{ 2 \mdsVariance }
+ \log  \normalCDF{ \frac{\modelDistance_{nn'}}{ \mdsSD} }
,
\label{eq:bmdsLikelihood}
\end{align}
for $\normalCDF{\cdot}$ the cumulative distribution function of a standard normal variate.  Unlike classical MDS, BMDS uses the language of probability to describe the low-dimensional representations, thus (1) exchanging the $O(N^3)$ computational complexity of classical MDS for the $O(N^2)$ complexity of evaluating the BMDS likelihood, (2) allowing for uncertainty quantification and (3) avoiding conceptual difficulties arising from the mixed-methodological application of probability models to the results of classical MDS. Indeed, in the BMDS framework, modeling the latent variables $\latentdata$ is as straightforward as specifying the prior distribution within a hierarchical model. Examples of such an approach are the use of a mixture of $D$-dimensional normals in \cite{oh2007model} and Gaussian processes in \cite{holbrook2021massive}, but there is no reason \emph{a priori} to restrict the class of available priors to be Gaussian.

\newcommand{\ra}[1]{\renewcommand{\arraystretch}{#1}}

\begin{figure}[!t]
	\centering
	\includegraphics[width=0.7\linewidth]{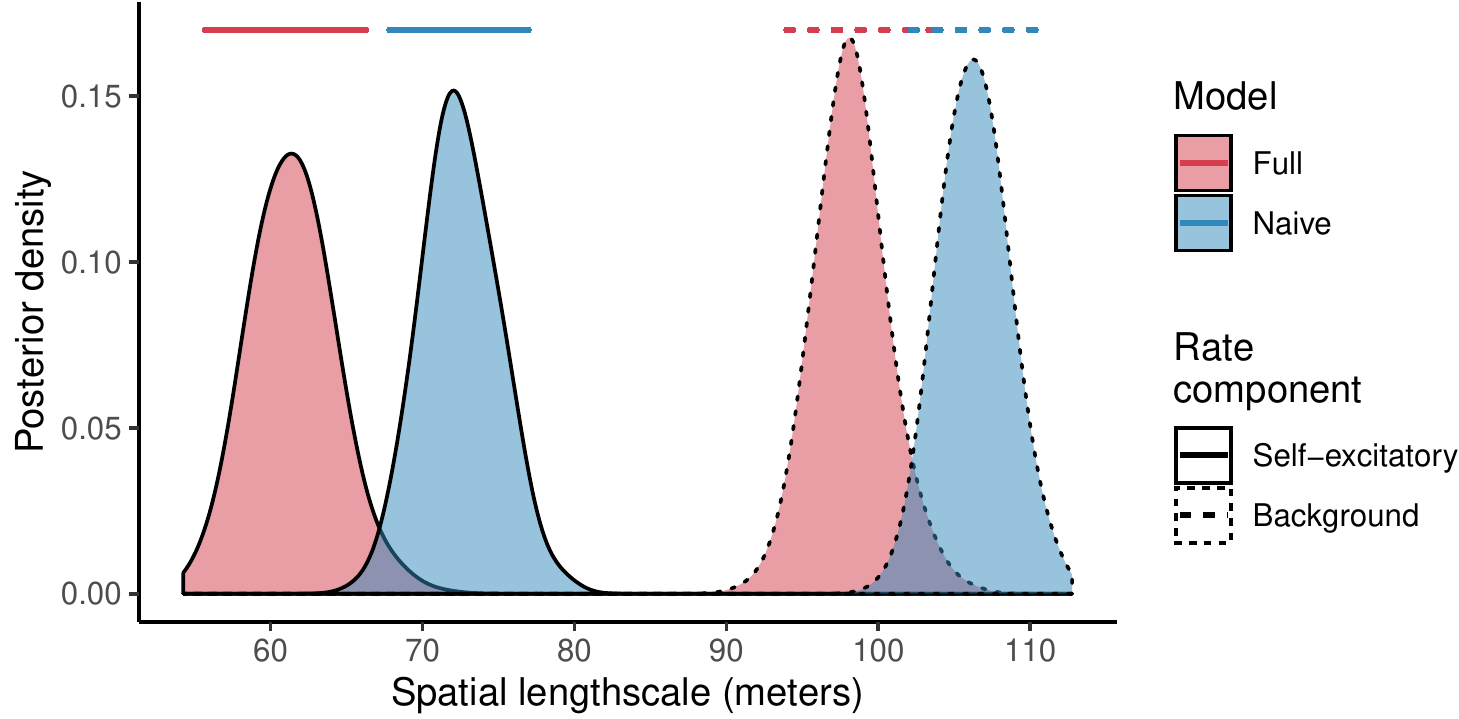} \\
	
	\vspace{1em}
	
	\ra{1.2}
	\resizebox{0.8\textwidth}{!}{\begin{tabular}{llll} 
			\toprule
			&& \multicolumn{2}{c}{Posterior median (95\% Credible interval)}  \\ \cmidrule{3-4}
			Rate component& Parameter & Full model & Naive model   \\
			\midrule
			Background  & Spatial lengthscale (m) & 98.1 (94.0, 103.3) & 106.3 (102.1, 110.7) \\
			& Temporal lengthscale (hrs) & 1763.7 (1552.9, 2014.8) & 1891.8 (1665.1, 2163.6) \\
			Self-excitatory & Spatial lengthscale (m) & 61.4 (56.4, 67.2) & 72.3 (67.9, 77.2) \\
			& Temporal lengthscale (hrs) & 0.009 (0.008, 0.010) & 0.009 (0.008, 0.009) \\
			& Normalized weight & 0.11 (0.10, 0.12) & 0.11 (0.10, 0.12) \\
			\bottomrule
	\end{tabular}}
	\captionlistentry[table]{A table beside a figure}\label{tab::gunshots}
	\captionsetup{labelformat=andtable}
	\caption{Posterior densities and 95\% credible intervals of Hawkes model background and self-excitatory lengthscales for `full' (locations inferred) and `naive' models of gun violence in the District of Columbia.   Credible intervals for self-excitatory lengthscales do not overlap, while those of the background component display marginal overlap.}\label{fig::spatLength}
\end{figure}

\paragraph{Choosing number of latent dimensions}

We would like to determine the optimal latent dimensionality $D$ for the spatiotemporal Hawkes process, we use $D$ to quantify the complexity of viral contagion through the global human air transportation network, and we let cross-validation \citep{geisser1975predictive} dictate our choice of $D$.
For BMDS, our data are the distance matrix $\mathbf{Y}$ with off-diagonal elements $\distance_{nn'}$ the pairwise distances between objects $n$ and $n'$.  Within $F$-fold cross-validation, each fold $f$ comprises held-out observations $\distanceMatrix^f$ and the remaining observations  $\distanceMatrix^{-f}$.  Let $s$ index an MCMC state corresponding to a single draw from the posterior conditioned on $\distanceMatrix^{-f}$ and denote the set of latent locations and model parameters $(\latentData,\Theta,\sigma^2)^{(s),-f}$ for $s=1,\dots,S$, the total number of MCMC states.  We take the empirical \emph{log pointwise predictive density} ($\widehat{lpd}$) as a measure of  model fit and start with the log pointwise predictive densitiy $lpd$ \citep{vehtari2017practical}:
\begin{align}\label{eq:lpd}
lpd &= \sum_f \sum_{n<n'} \log p(\distance_{nn'}^f|\distanceMatrix^{-f})  \\ \nonumber
&= \sum_f  \sum_{n<n'} \log \int p(\distance^f_{nn'}|\latentData,\Theta,\sigma^2) p(\latentData,\Theta,\sigma^2|\distanceMatrix^{-f}) \, \dx (\latentData,\Theta,\sigma^2) \\ \nonumber
&\approx \sum_f  \sum_{n<n'} \log \frac{1}{S} \sum_{s=1}^{S} p(\distance^f_{nn'}|(\latentData,\Theta,\sigma^2)^{(s),-f}) = \widehat{lpd} \, .
\end{align}
Given competing models with different latent dimensionalities, we generally prefer  the model with larger $\widehat{lpd}$.

\section{Inference and implementation}\label{sec:inference}

We approach all three applications with an adaptive random scan Metropolis-within-Gibbs \citep{gilks1995adaptive} scheme building on Algorithm 1 of \citet{holbrook2021scalable}. For the first two applications, the target posterior distribution takes the form
\begin{align*}
\cdensity{\Theta}{\mathfrak{X}, \ttimes}
\propto
\cdensity{\mathfrak{X},\ttimes}{ \Theta}
\, \density{\Theta}
=
\left(
\int
\cdensity{\mathfrak{X}}{\latentData}
\mathcal{L}(\latentData,\ttimes|\, \Theta)\,
\dx \latentData
\right)
\density{\Theta} \, ,
\end{align*}
where one obtains the uniform $\cdensity{\mathfrak{X}}{\latentData}$ by inverting the constraints of Equations \eqref{eq:locsPrior1} and \eqref{eq:locsPrior2}.  We compute the high-dimensional integral over $\latentData$ using a Metropolis-Hastings kernel with block-wise updates over sets of individual location variables.  Satisfying the square constraints of Equation \eqref{eq:locsPrior1} is straightforward using truncated normal proposals.   Satisfying the circular constraints of Equation \eqref{eq:locsPrior2} is less cut and dry.  For an individual location variable $\x_n$ of the Markov chain's state $s$, we generate the $(s+1)$th state according to the Metropolis kernel with proposal distribution
\begin{align}\label{eq:circleKernel}
\x_n^* \sim q(\x_n^*|\x_n)  \propto 1 \, , \quad   ||x_n^*-\mathfrak{x}_n||_2 < r_n \,,  \quad ||x_n^*-\x_n||_2 < r_n\epsilon\, , \quad \epsilon >0 \, ,
\end{align}
where $\epsilon$ is an algorithmic parameter tuned with the help of diminishing adaptations \citep{roberts2007coupling}. When $||\x_n-\mathfrak{x}_n||_2 + r_n\epsilon < r_n$, sampling is easy.  Otherwise, we use a simple rejection sampler that satisfies the two circular constraints.  In this case, the Metropolis-Hastings accept-reject step requires calculating the area of the two circles' intersection (also referred to as the asymmetric lens), but one can easily obtain this quantity in closed-form and with negligible computational expense.
Under our BMDS formulation for the third application, global viral contagion, the target posterior distribution is
\begin{align*}
\cdensity{\mdsVariance, \Theta}{\traitData, \ttimes}
\propto
\cdensity{\traitData}{ \mdsVariance}\,
\density{\mdsVariance}\,
\density{\Theta}
=
\left(
\int
\cdensity{\traitData}{\latentData, \mdsVariance}
\cdensity{\latentData,\ttimes}{\Theta}\,
\dx \latentData
\right)
\density{\mdsVariance}\,
\density{\Theta} \, .
\end{align*}
To compute the high-dimensional integral over values of $\latentData$, we use Hamiltonian Monte Carlo (HMC) \citep{neal2011mcmc} and again use simple Metropolis-Hastings proposals for the remaining parameters $\mdsVariance$ and $\Theta$.   Hamiltonian Monte Carlo  over $\latentData$ requires evaluation of both spatiotemporal Hawkes model and BMDS joint densities (Equations \ref{eq:likelihood} and \ref{eq:bmdsLikelihood}) and their gradients.

The Hawkes model likelihood (used in all three applications) and the BMDS likelihood and Hawkes likelihood gradients (used in the third application)  all share the prohibitively burdensome computational complexity $O(N^2)$. We therefore use the \textsc{OpenCL} and \textsc{C++}  high performance computing  libraries \textsc{MassiveMDS} \citep{holbrook2021massive} \url{https://github.com/suchard-group/MassiveMDS} and \textsc{hpHawkes} \citep{holbrook2021scalable} \url{https://github.com/suchard-group/hawkes} to evaluate these functions and their gradients in parallel on either a graphics processing unit (GPU) or with a multi-core central processing unit (CPU) with vectorization.  In writing this paper, we have contributed GPU and CPU implementations of the Hawkes process log-likelihood gradient to the library \textsc{hpHawkes}, and we detail the massively parallel Algorithms \ref{alg:lik2} and \ref{alg:lik} and their resulting speedups in Section \ref{sec:parallelization}.  Finally, we access and embed the high performance implementations within the broader Metropolis-within-Gibbs scheme with the \textsc{BEAST} software package \citep{suchard2018bayesian} using simple application programming interfaces.

\section{Demonstrations}

\begin{figure}[!t]
	\centering
	\includegraphics[width=0.7\linewidth]{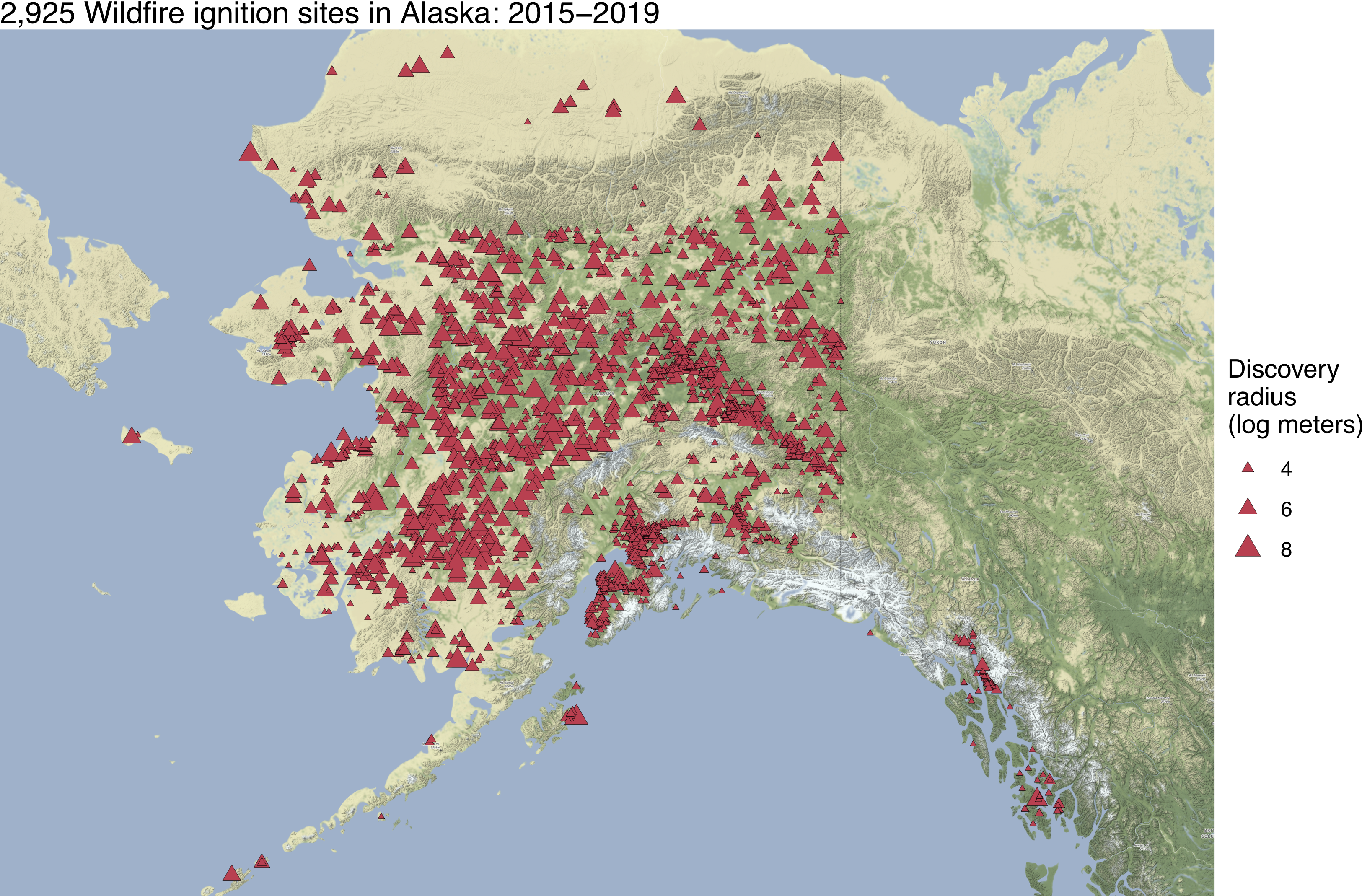}
	\caption{Data include time of fire discovery as well as size (in acres) and location of each wildfire at time of discovery.}\label{fig:alaskamap}
\end{figure}

Besides the high-performance computing packages \textsc{hpHawkes} and \textsc{MassiveMDS} we use for MCMC, we use the \textsc{R} programming language \citep{rlang} and the \textsc{R} graphics packages \textsc{ggplot2} \citep{ggplot} and \textsc{ggmap} \citep{kahle2013ggmap} to produce and summarize results.  The \textsc{R} package \textsc{coda} \citep{coda} provides our effective sample size (ESS) measures, and we base reported 95\% credible intervals on empirical posterior 0.025 and 0.975 quantiles.  Finally, we make all analysis source files publicly available at \url{https://github.com/andrewjholbrook/unknown_locs} and \url{https://github.com/andrewjholbrook/FluHawkes}.

\subsection{Washington D.C.~gunfire in 2018}

We first apply our methodology to the analysis of 3,982 gunshots occurring in Washington D.C. between January 2 and December 31, 2018.  The Government of the District of Columbia makes gunfire data from the years 2014-2019 freely available at \url{https://opendata.dc.gov}.  The data arise from ShotSpotter AGLS technology \citep{carr2016geography} that has become increasingly accurate since first implemented in Washington D.C.~in 2006.  \citet{loeffler2018gun} use a spatiotemporal Hawkes process to analyze a similar sample from the years 2010 through 2012, and \citet{holbrook2021scalable} apply a related model to data from the years 2006 through 2019. The D.C.~Government rounds all latitudinal and longitudinal coordinate data to three decimal places, a coarsening that corresponds to 100 meters precision. Because we wish to isolate this as the only source of spatial uncertainty and because of the gradual improvement of ShotSpotter technology, we choose to focus on a higher quality sample from 2018.  We also remove all observations listed as potential firecrackers as well as all data from the first day of the year, again avoiding possible corruption due to misattribution to firecrackers.  The result is 3,982 events with locations plotted in Figure \ref{fig:dcmap}. The minimum, mean and maximum pairwise distances between raw locations are 0.0, 5.4 and 16.4 km.  We compare these numbers to the data's spatial precision of 0.1 km.

To infer all six Hawkes model parameters and all 3,982 location variables, we generate 30 million Markov chain states (requiring 48 hours on our Nvidia Quadro GP100 GPU) that provide minimum and mean ESS of 131 and 424 for latent locations and 401 and 571 for model parameters.  First, we would like to know whether there is a spatial pattern to the posterior displacements away from raw locations for individual events, where we use the formula $||\sum_{s=1}^S (\x_n^{(s)}-\mathfrak{x}_n )/S||_2$ to quantify this displacement.  Figure \ref{fig:dcmap} shows that high posterior displacements occur both on the peripheries and at the centers of high activity areas.  This fact suggests that more complex patterns underlie posterior displacements and that temporal relationships may play a role.  Figure \ref{fig:dcmapZoom} presents three examples of event groups with larger posterior displacements.  For the first group, posterior locations draw away from figure center despite relatively small temporal differences between events ranging from 11 to 120 hours.  Here, temporal proximity appears to be overcome by the gunfire vacuum of a commercial shopping center at plot center.  The second pair of events appear to have larger posterior displacements for a very different reason.  Here, the two events are spatially isolated from other gunshots, so their posterior locations attract to each other, despite a larger temporal disparity of 55 days.  Finally, the third cluster tells a much simpler story.  The four events occupy the center of a large, high-activity area and gravitate toward the center of mass.

Figure \ref{fig::spatLength} and Table \ref{tab::gunshots} present inferential results for the Hawkes model parameters, where the normalized self-excitatory weight $\theta/(\theta+\mu_0)$ communicates the proportion of all events arising from self-excitation rather than the background process.  Here, we see generally consistent results between the model that incorporates spatial uncertainty and that which does not.  Perhaps unsurprisingly, the major discrepancies in posterior inference between these models are for the two spatial lengthscales.  The self-excitatory spatial lengthscales for the full and naive models are 61.4 m (56.4, 67.2) and 72.3 m (67.9, 77.2), and the background spatial lengthscales are 98.1 m (94.0, 103.3) and 106.3 m (102.1, 110.7).  Smaller spatial lengthscales make sense insofar as inferred locations may attract to each other, but one might find these statistically significant and marginally statistically significant differences surprising given the relatively small spatial uncertainty (0.1 km) precision relative to a mean pairwise distance in the data of roughly 5 km.  Despite statistically significant differences, we judge the practical differences to be small, and this is good news for practitioners who want to avoid integrating over latent locations.  Nonetheless, this good news only seems to apply when spatial uncertainty is (a) relatively small and (b) uniform across observations \emph{a priori}.  In Section \ref{sec:sim}, we use simulated data to test this hypothesis and find that the naive model indeed fails under moderate coarsening.

\begin{figure}[t!]
	\centering
	
	\includegraphics[width=0.8\linewidth]{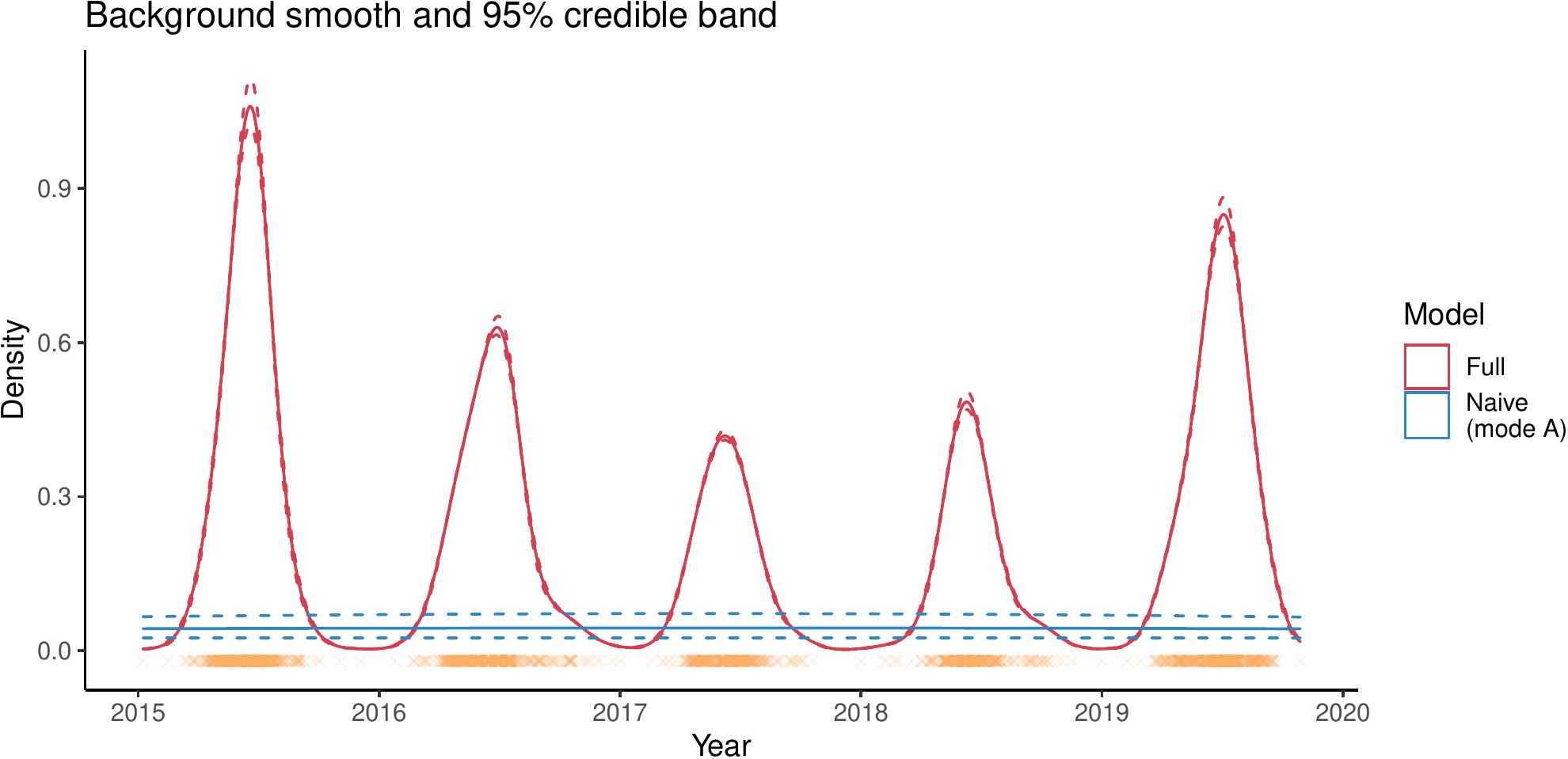}
	
	\vspace{1em}
	
	\resizebox{0.9\textwidth}{!}{\begin{tabular}{lllll} 
			\toprule
			&& \multicolumn{3}{c}{Posterior median (95\% Credible interval)}  \\ \cmidrule{3-5}
			Rate component& Parameter & Full model & Naive model \emph{A} & Naive model \emph{B}  \\
			\midrule
			Background  & Spatial lengthscale (km) & 34.8 (32.9, 37.6) &  \textbf{23.5 (22.3, 24.6)} & 63.0 (58.7, 68.7) \\
			& Temporal lengthscale (days) & 25.9 (23.8, 27.9) & 3244.0 (1929.7, 5803.5)& \textbf{10.2 (9.4, 11.1)}\\
			Self-excitatory & Spatial lengthscale (km) & 11.1 (10.1, 12.0) & \textbf{23.3 (22.2,   24.4)} & 6.5 (5.9, 7.2)\\
			& Temporal lengthscale (days) & 1.1 (0.9, 1.4)  & 2.2 (1.9, 2.5) & \textbf{10.0 (9.2, 10.8)} \\
			& Normalized weight & 0.34 (0.31, 0.37)  & 0.27 (0.17, 0.36) & 0.44 (0.41,  0.47)\\
			\bottomrule
	\end{tabular}}
	\captionlistentry[table]{A table beside a figure}\label{tab::wildfires}
	\captionsetup{labelformat=andtable}
	\caption{Spatiotemporal Hawkes model posterior inference for 2015-2019 wildfire ignitions in Alaska with `Full' model (locations inferred) and of `Naive' model (locations not inferred) modes \emph{A} and \emph{B}: inferring locations may help avoid modes at near-equal lengthscales (\textbf{\emph{bold}}). Mode \emph{A} provides an unreasonably large background temporal lengthscale that fails to incorporate seasonal trends, and the normalized self-excitatory weight of mode \emph{B} may be considered too large to be realistic.}\label{fig::tempLength}
\end{figure}

\subsection{Alaskan wildfire ignitions: 2015-2019}

The Alaska Interagency Coordination Center makes various wildfire data resources freely accessible at \url{https://fire.ak.blm.gov/predsvcs/maps.php}.  In particular, we apply our methodology to data consisting of wildfire geographic coordinates, date and time of fire discovery and size in acres at time of discovery.  Figure \ref{fig:alaskamap} displays the raw locations for all 2,925 wildfires, plotting each with size proportional to its radius on the log scale.  The minimum, mean and maximum pairwise distances between raw locations are 0.0, 500.3 and 2,373.8 km.  The empirical distribution of wildfire discovery site radii roughly resembles a power law, with minimum, median, mean and maximum of 0.01, 0.01, 0.08 and 4.42 km.  In this way, the spatial uncertainty relative to the scale of locational spread is much smaller for this application than for the Washington D.C.~gunfire example.  This difference allows us to partially isolate the effects of differential uncertainty across the observed sample on posterior inference.

Using the prior of Equation \ref{eq:locsPrior2} on locations, we model the spread of wildfire ignitions with our spatiotemporal Hawkes model and simultaneously infer ignition locations.  Again, we generate 30 million Markov chain states that provide a minimum and mean ESS of 120 and 398 for inferred locations and 510 and 586 for the six Hawkes model parameters.  Despite the smaller sample size compared to the gun violence data, MCMC for this example requires 50\% more time (totaling roughly 72 hours) due to the rejection sampler we use to generate from \eqref{eq:circleKernel}.
Figure \ref{fig::tempLength} and Table \ref{tab::wildfires} present inferential results from the model with locations inferred as well as the naive model.  As discussed in Section \ref{sec:hawkesModel}, simultaneous inference for self-excitatory and background lengthscales can sometimes lead to multimodality. This issue is so problematic, that \citet{mohler2014marked} fixes the problem by setting spatial lengthscales to be equal, i.e., by removing flexibility from the model.  Here, we find that inferring locations may actually help mitigate such multimodality: while the naive model gets stuck in two different modes ($A$ and $B$), the full model does not.  As shown in Figure \ref{fig::tempLength}, multimodality can lead to poor model fits and unreasonable results.  With a posterior mean of 3,244.0 days and 95\% credible interval of  (1,929.7, 5,803.5), inference for the background temporal lengthscale of the naive model's mode A suggests no seasonal wildfire trend whatsoever \emph{in Alaska}.  In contrast, inference for the full model fully captures seasonal trends with a posterior mean of 25.9 days (23.8, 27.9) for the temporal lengthscale of the background rate.

\begin{figure}[tp!]
	\centering
	\includegraphics[width=0.9\linewidth]{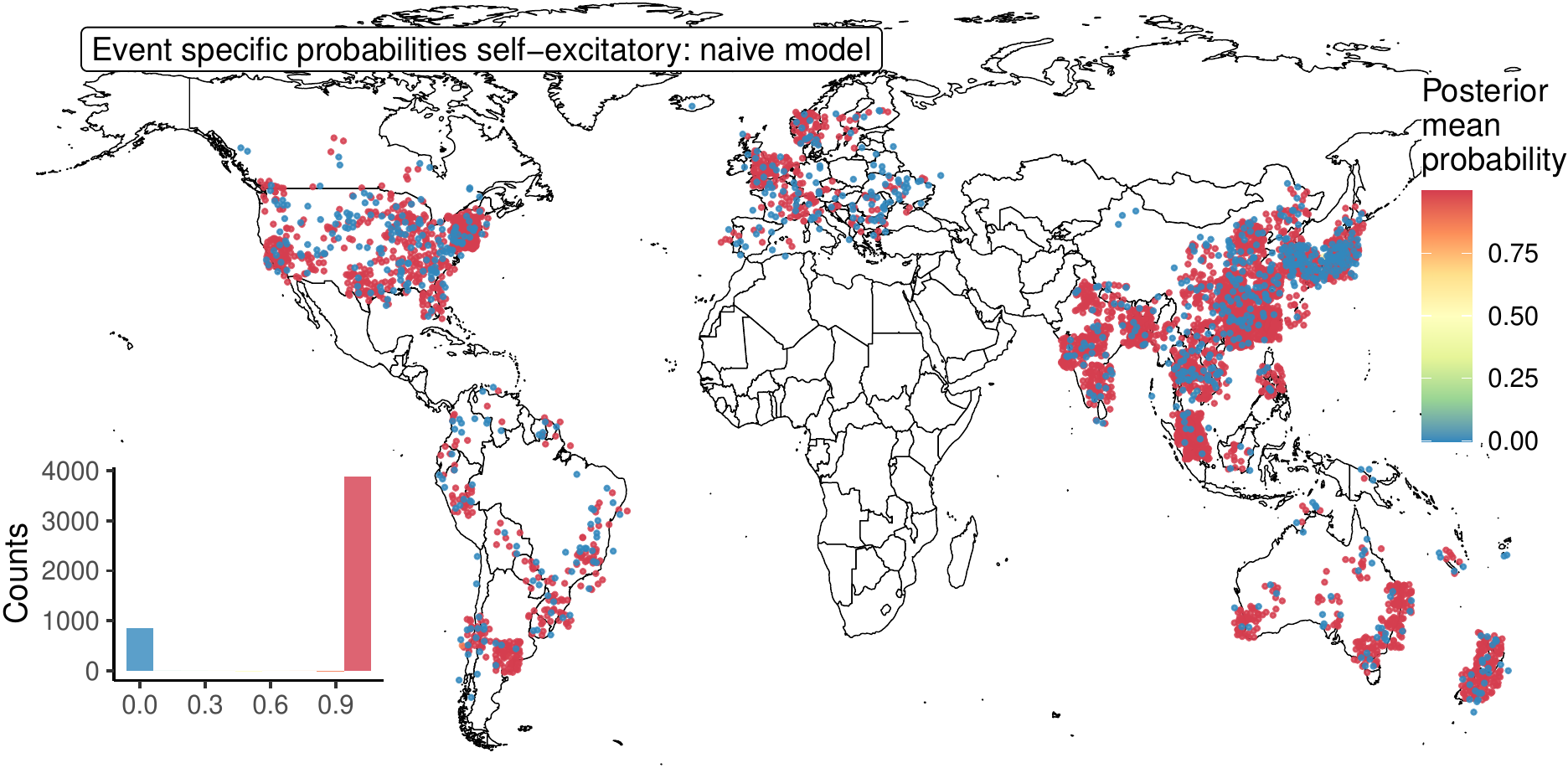}
	\includegraphics[width=0.9\linewidth]{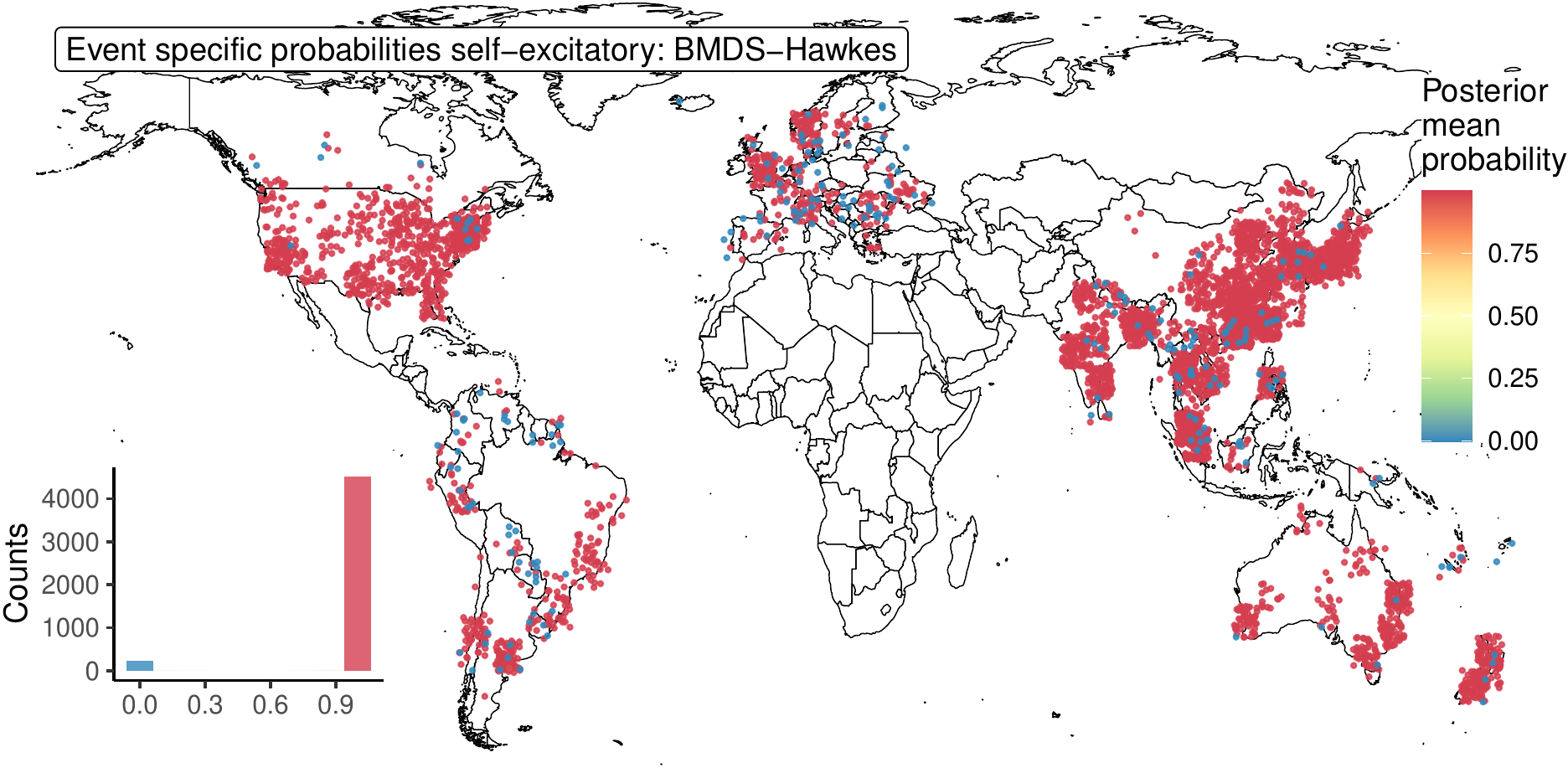}
	\includegraphics[width=0.9\linewidth]{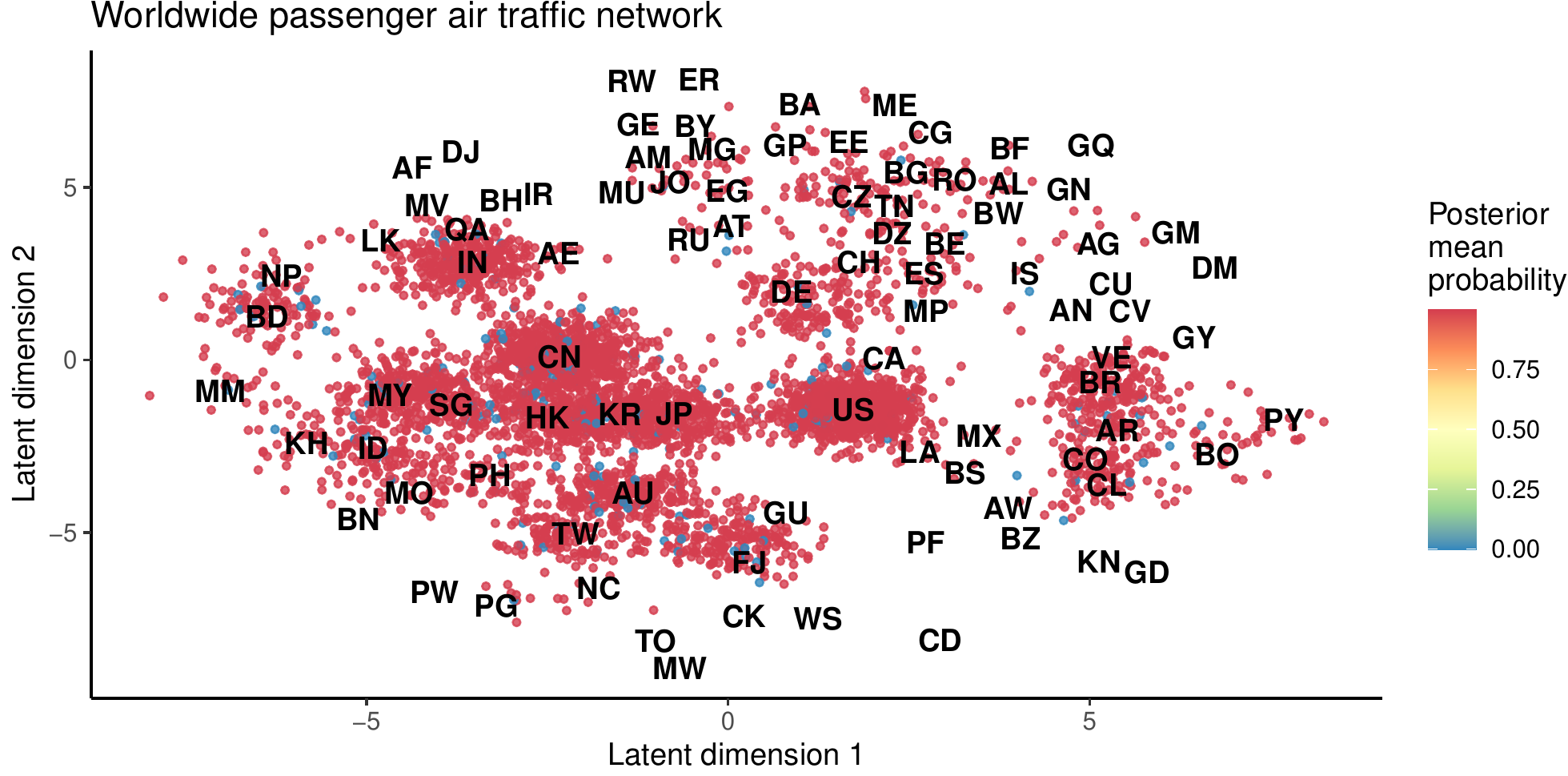}
	\caption{Geographic and network positions of 4,733 influenza cases, each colored by posterior mean probability the case originates from another `parent' case.  Top figure shows results from spatially naive model; bottom two figures from the  6- and 2-dimensional combined Bayesian multidimensional scaling and spatiotemporal Hawkes model (BMDS-Hawkes). In the latter, proximity of Laos (LA) to the United States (US) portends poor cross-validation results.}\label{fig:globalMaps}
\end{figure}

\subsection{Global influenza cases: 2000-2012}\label{sec:bmdsHawkesDemo}

We analyze the worldwide spread influenza using 4,733 cases recorded between the years of 2000 and 2012.  Of the 4,733 cases, 1,161 are H1N1 subtype, 1,341 H3N2 subtype, 1,195 Victoria lineage (VIC) and 1,036 Yamagata lineage (YAM).  H1N1 and H3N2 are influenza type A and generally more prevalent than Victoria and Yamagata, which are both type B and contribute to significantly less infections annually.  Between H1N1 and H3N2, H1N1 is responsible for two major pandemics, the Spanish flu of 1918-1919 and the swine flu of 2009, while H3N2 has contributed to one, the Hong Kong flu of 1968-1969. \cite{bedford2015global} relate the greater epidemiological success of type A influenza to higher rates of antigenic drift, leading to different age groups becoming infected at different rates.  In particular, adults are more susceptible to H1N1 and, being more likely frequent fliers than children, help the subtype travel more quickly through global air travel networks \citep{bedford2014integrating} than competing strains. Combining BMDS with a phylogenetic diffusion model that conditions on each subtype and lineage's evolutionary history, \citet{holbrook2021massive} confirms that the rate of diffusion through the global air traffic network is significantly greater for H1N1 than for H3N2, YAM and VIC.  Here, we are interested in whether inference based on our BMDS-Hawkes model renders similar results and how greater efficiency of H1N1 might express itself for individual Hawkes model parameters, e.g. shorter lengthscales or greater rates of self-excitation.

The data we consider here are a subset of the 5,392 analyzed in \citet{holbrook2021massive}, where we have removed those cases that lack a precise date.  Moreover, we use the exact same matrix of pairwise air traffic distances between countries $\traitData$.  \citet{brockmann2013hidden} creates these distances from a network for which nodes are 4,096 airports worldwide and edges (when they exist) between nodes inversely relate to the total number of passengers traveling between the two airports each year. Motivated by the multi-precision nature of the influenza case data---spatial labels are approximately 1/3 cities, 1/3 provinces and 1/3 countries---\citet{holbrook2021massive} then collapse across airports to obtain effective distances between countries on this global transportation network.  We use the Hawkes model to infill the relationships between latent locations coming from the same country. Through the spatiotemporal Hawkes likelihood that interfaces with temporal data $\ttimes$, the BMDS-Hawkes model further informs latent positions $\latentData$.  Thus, we efficiently and simultaneously (a) adapt our data to the realities of global air transport and (b) exploit all data despite its multi-precision nature.

Before producing the full analysis, we use the $\widehat{lpd}$ of Equation \eqref{eq:lpd} as measure of model fit and perform 5-fold cross-validation to select the latent dimensionality of our BMDS-Hawkes model.  Dimensions 2 through 8 provide $\widehat{lpd}$s of -13.2, -8.1, -6.2, -5.5, -5.2, -5.1 and -5.0 million.  Noting a lack of relative improvement for further dimensions, we judge the 6-dimensional model to be sufficiently complex.   Next, we use HMC (Section \ref{sec:inference}) to generate 80 million Markov chain states. Employing Algorithm \ref{alg:lik} for massively parallel Hawkes log-likelihood gradient calculations, this requires roughly 10 days on our Nvidia Titan V GPU.

The top two plots of Figure \ref{fig:globalMaps} show the naive global distribution of the influenza case data colored by the posterior mean probability that each event arises from self-excitation, i.e.,
\begin{align*}
\frac{1}{S} \sum_{s=1}^S \xi^{(s)}(\x_n^{(s)},t_n)/ \left(\xi^{(s)}(\x^{(s)}_n,t_n)+ \mu^{(s)}(\x^{(s)}_n,t_n)\right)
\end{align*}
for $\x^{(s)}_n$ a location in the 6-dimensional latent air traffic network space and $\xi^{(s)}(\cdot,\cdot)$ and $\mu^{(s)}(\cdot,\cdot)$ the self-excitatory and background rates parameterized by parameters $\Theta^{(s)}$.  Posterior concentration around this posterior mean is extremely tight for all observations, so the models strongly believe that blue cases arise from the background process while red arise from self-excitation.  For the naive model, there are as many blue cases as there are locations: the model regards the earliest case in every location as coming from the background process and every case thereafter as arising from this earliest case. Reflecting this fact, the posterior distributions of the naive model's self-excitatory spatial lengthscale concentrate below 3 km (Figure \ref{fig:naiveInf}).  Thus, naive model inference communicates no information beyond what one would garner from a simple exploratory data analysis.  On the other hand, the 6-dimensional BMDS-Hawkes model reveals significantly less background activity and a model more in tune with the self-excitatory reality of viral spread. Still, we can interpret background activity as arising from relatively large and fast traversals of the global air traffic network.
The third plot is similar, but shows the arrangement of latent locations for a single posterior sample for the 2-dimensional BMDS-Hawkes model.  In general, the world economic powers gravitate toward the middle while smaller countries tend toward the outside.  These arrangements are largely as one might hope, but there are hints that the 2-dimensional model is insufficient.  For example, Laos (LA) is much closer to the United States (US) than it is the rest of Asia in general and China in particular.  This suggests that a higher dimensionality might be more appropriate, a fact which cross-validation results bear out.

\begin{figure}[t!]
	\includegraphics[width=0.9\linewidth]{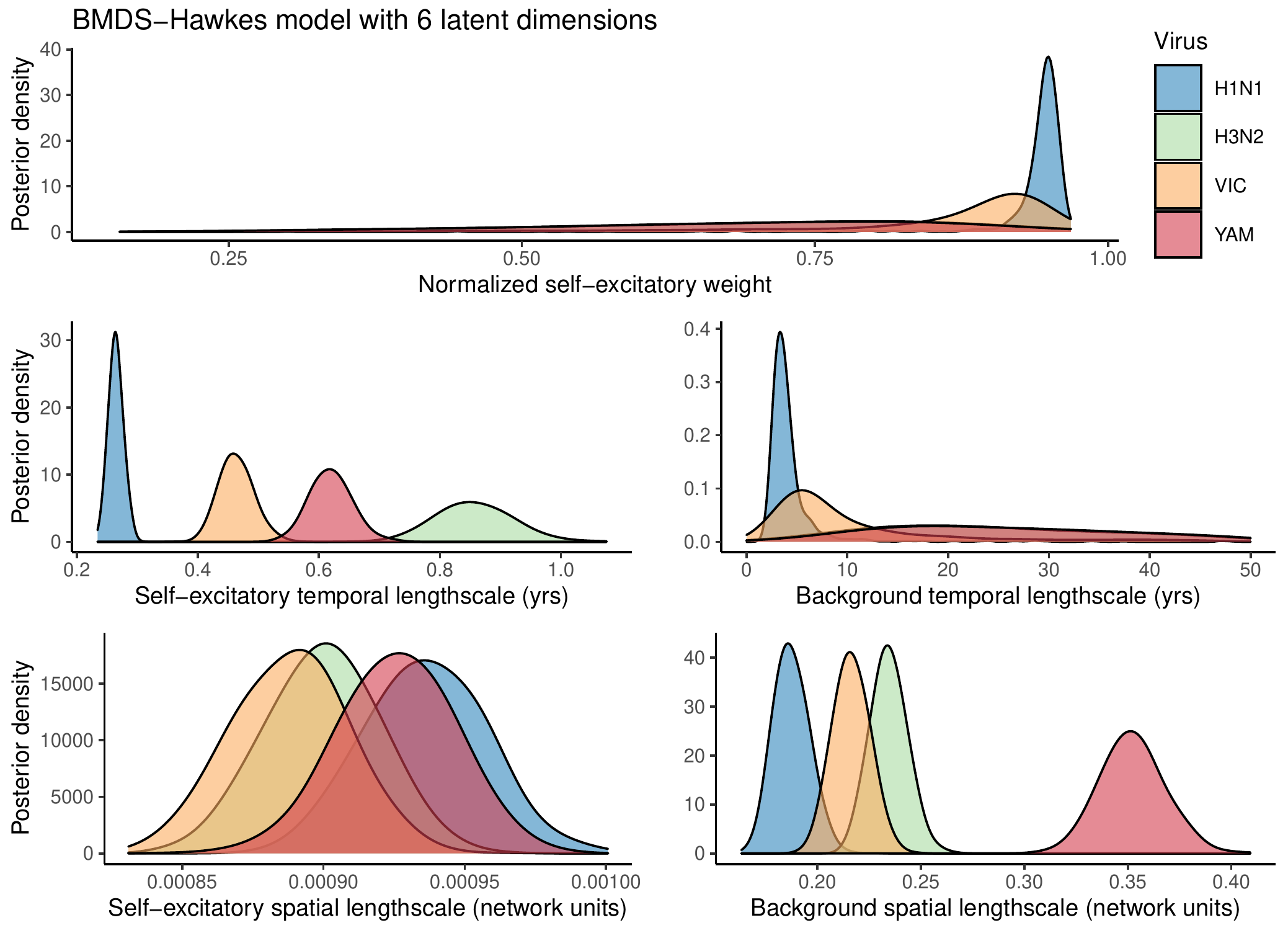}
	\caption{Posterior distributions for Hawkes model parameters based on 1,161 H1N1 subtype, 1,341 H3N2 subtype, 1,195 Victoria lineage and 1,036 Yamagata lineage influenza cases. In general, H1N1 is more prevalent and infects adults in greater numbers than it does children.}\label{fig:fullInf}
\end{figure}

Figure \ref{fig:fullInf} displays posterior densities for the Hawkes model parameters for each influenza subtype and lineage.  We  immediately notice that the H1N1 model attributes more influenza activity to self-excitation than do the other models.  The posterior means and 95\% credible intervals of the normalized self-excitatory weight for H1N1, H3N2, VIC and YAM are 0.95 (0.82, 0.96), 0.72 (0.31, 0.91), 0.91 (0.47, 0.95), 0.72 (0.32, 0.90).  We note that overlapping credible intervals indicate uncertainty as to this specific ordering.  On the other hand, there is very little posterior uncertainty with regards to the ordering of the self-excitatory temporal lengthscales. In order, the same posterior measures for H1N1, VIC, YAM and H3N2 are 0.26 years (0.24, 0.29), 0.46 years (0.42, 0.52), 0.62 years (0.56, 0.69) and 0.86 years (0.74, 0.98).  This result suggests that the self-excitatory temporal lengthscale of our Hawkes model and the rate of diffusion of the phylogenetic diffusion model of \citet{holbrook2021massive} are similar insofar as they both capture the greater efficiency with which H1N1 uses passenger air traffic networks to quickly travel the globe.  Moreover, H1N1's posterior mean self-excitatory temporal lengthscale is small enough to fully capture seasonal trends.  Finally, the large posterior mean for the same parameter of the H3N2 model and its inability to capture seasonal trends suggests that a network build solely of air transportation may be insufficient for modeling the spread of H3N2.

\section{Discussion}

The spatiotemporal Hawkes process is a powerful tool for modeling the complex spatial dynamics of many real world phenomena. Although its use is growing increasingly widespread, previous applications of the model have glossed over the presence of spatial coarsening and uncertainty in the problems analyzed and the role played by the same uncertainty in biasing model inference.   By considering three diverse applications, we have demonstrated (a) the prevalence of spatial uncertainty in processes commonly regarded as self-exciting, (b) the practicality of integrating over such uncertainty in the manners proposed and (c) the statistically and practically significant differences between full and naive approaches.  Furthermore, we have shown that our strategies may also be useful in mitigating multimodality, a problem that makes model fitting, diagnostics and interpretation more cumbersome. Indeed, we have demonstrated that one can reap these benefits without having to sacrifice model flexibility.

That said, there are a few meaningful changes to our proposed approach that may lead to improved inference.  First, this approach seeks to account for spatial coarsening but fails to adapt to temporal coarsening.  For this reason, we removed hundreds of viral cases that lack full temporal precision from our analysis of global influenza.  One could make retainment of these observations possible by directly modeling the coarsened data in a similar way to how we have modeled locations in our first example.  Unfortunately, implementing the MCMC to integrate over latent times would be especially difficult due to the self-excitatory rate function's reliance only on past events.  The required combinatorial integration would necessitate careful bookkeeping, and it is difficult to predict the empirical mixing of such a Markov chain.  Second, our priors over locations in both the gunfire and wildfire examples could take further advantage of geographical information.  In the former example, one might use the fact that gun violence is more likely to occur indoors.  In the latter, one could use rivers and lakes to further reduce the support of wildfire ignitions.  We are not sure how one might generate and respect such priors at non-trivial scale. Third, we use Bayesian multidimensional scaling to model viral spread through Euclidean space instead of a complex network. It is plausible that other continuous spaces might be more appropriate. For example, a low-dimensional torus with its varying curvature might be better able to capture pairwise relationships more efficiently than Euclidean space of the same dimension.

There are usually many valid ways to approach a challenge, and other approaches may prove more tractable over time.  In modeling global viral spread, for instance, it might also prove useful to equip the self-excitatory rate with a dramatically heavy-tailed triggering function such as a continuous scale-mixture of truncated Gaussian functions \citep{polson2014bayesian,nishimura2018prior}.  This would allow the same triggering function to account for both domestic and international transmission. That said, this would ameliorate the issue of adapting to transportation dynamics but would not account for spatially coarsened and multi-precision data.

To conclude, we hope that this paper's results will prove instructive to scientists interested in spatiotemporal modeling regardless of the model or statistical paradigm they choose to employ.  At the very least, our results suggest that sensitivity testing by perturbing spatial data is a good idea.  In an exact analogy to the way we have selected priors on spatial locations, one might generate perturbations in a way that reflects the a reasonable model of the spatial uncertainty at hand.  Finally, our proposed approach appears to be easily translatable into the frequentist paradigm in the form of marginal maximum likelihood \citep{geyer1991markov}.  MCMC similar to that performed here would provide for integration over locations, although it is not immediately clear how such integration would influence the already complex consistency arguments involved in maximum likelihood estimation for spatiotemporal model parameters \citep{schoenberg2016note}.

\begin{appendix}
\section{Parallelization}\label{sec:parallelization}

In writing this paper, we have developed massively parallel implementations of the gradient of the log-likelihood with respect to spatial locations for our spatiotemporal Hawkes model.  We have also added this code to \textsc{hpHawkes}, a \textsc{C++} library and \textsc{R} package for high-performance computing for Bayesian inference under the spatiotemporal Hawkes process.  We have made this open-source software freely available at \url{https://github.com/suchard-group/hawkes}.

\begin{figure}[t!]
	\includegraphics[width=\linewidth]{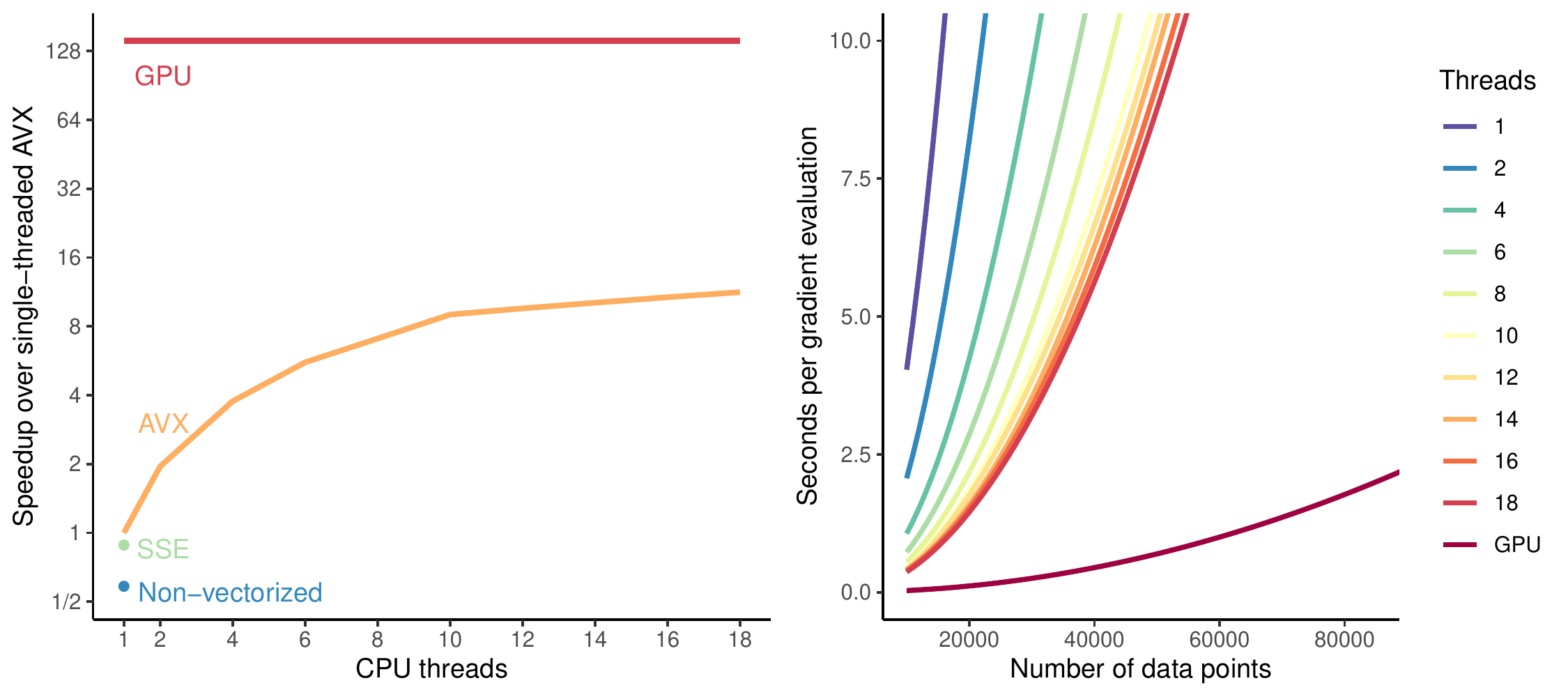}
	\caption{Computing the spatiotemporal Hawkes process log-likelihood gradient with respect to locations using central and graphics processing units (CPU and GPU). [Left] Relative speedups over single-core advanced vector extensions (AVX) vectorization for single-core non-vectorized and streaming SIMD extensions (SSE), multi-core AVX and many-core GPU implementations for 75,000 simulated observations. [Right] Absolute time to perform gradient evaluation for single- and multi-core AVX processing and GPU as a function of the number of simulated data points.}\label{fig:speedups}
\end{figure}

Letting $\lambda_{nn'}$ be as defined in Equation \eqref{eq:likelihood}, define $\lambda_n := \sum_{n'}^N \lambda_{nn'}$.  Then the gradient of the Hawkes likelihood with respect to locations is
\begin{align*}
\frac{\partial \ell}{\partial \latentdata_n} &= \sum_{n'=1}^N \left( \frac{\mu_{nn'}}{\lambda_{n}} + \frac{\mu_{n'n}}{\lambda_{n'}}\right) \left(\frac{\latentdata_{n'}-\latentdata_{n}}{\tau_x^2} \right) +  \left( \frac{\xi_{nn'}}{\lambda_{n}} + \frac{\xi_{n'n}}{\lambda_{n'}}\right) \left(\frac{\latentdata_{n'}-\latentdata_{n}}{\sigma_x^2} \right)  \\
&:=  \sum_{n'=1}^N \left(\frac{\partial \ell}{\partial \latentdata_n} \right)_{n'} \, .
\end{align*}
Algorithms \ref{alg:lik2} and \ref{alg:lik} describe our massively parallel implementation of this gradient on a CPU and GPU, respectively, and Figure \ref{fig:speedups} illustrates some of the speedups achieved by these implementations.

\newcommand{\bb}{\mathbf{b}}
\newcommand{\vv}{\mathbf{v}}
\newcommand{\aaa}{\mathbf{a}}
\newcommand{\adapt}{\mathbf{l}}

\newcounter{algsubstate}
\renewcommand{\thealgsubstate}{\alph{algsubstate}}
\newenvironment{algsubstates}
{\setcounter{algsubstate}{0}%
	\renewcommand{\State}{%
		\stepcounter{algsubstate}%
		\Statex {\footnotesize\thealgsubstate:}\space}}
{}

\newcommand{\transformR}{r}
\newcommand{\transformCDF}{c}
\newcommand{\threadsPerBlock}{B}

\algblock{ParFor}{EndParFor}
\algnewcommand\algorithmicparfor{\textbf{parfor}}
\algnewcommand\algorithmicpardo{\textbf{do}}
\algnewcommand\algorithmicendparfor{\textbf{end\ parfor}}
\algrenewtext{ParFor}[1]{\algorithmicparfor\ #1\ \algorithmicpardo}
\algrenewtext{EndParFor}{\algorithmicendparfor}

\newcommand{\blockSize}{B}

\newcommand{\llambda}{\boldsymbol{\lambda}}
\newcommand{\Ddelta}{\boldsymbol{\Delta}}

\begin{algorithm}
	\scriptsize
	\caption{Parallel computation of Hawkes process log-likelihood gradient: 
		\emph{uses multiple central processing unit (CPU) cores along with loop vectorization to compute log-likelihood gradient.  For double-precision floating point, the algorithm uses either SSE or AVX vectorization to make $J=2$ or $4$ long jumps and cut loop iterations by one-half or three-fourths, respectively. Here, $B$ is the number of CPU threads available.   Symbols $\ell$, $\lambda$ and $\Lambda$ appear in Equation \eqref{eq:likelihood}.}}\label{alg:lik2}
	\begin{algorithmic}[1]
		\State Compute rates $\lambda_1, \dots, \lambda_N$:
		\begin{algsubstates}
			\State \hspace{1em} \textbf{parfor} $b \in \{1,\dots,B\}$ \textbf{do}
			\State \hspace{2em} \textbf{if} $b\neq B$ \textbf{then}
			\State \hspace{3em}$Upper \gets b \lfloor  N/B  \rfloor$
			\State \hspace{2em} \textbf{else}
			\State \hspace{3em} $Upper \gets  \lceil  N/B  \rceil$
			\State \hspace{2em} \textbf{end if}
			\State \hspace{2em} \textbf{for} $n \in  \{ (b-1)\lfloor N/B  \rfloor + 1,  \dots,Upper \}$ \textbf{do}
			\State \hspace{3em} copy $\x_{n}$, $t_{n}$ to cache
			\State \hspace{3em} $\llambda_{n} \gets \mathbf{0}$ \Comment{vector of length J}
			\State  \hspace{3em} $n' \gets 1$
			\State  \hspace{3em} \textbf{while} $n' < N$ \textbf{do}
			\State \hspace{4em} $J \gets \min(J,N-n')$
			\State \hspace{4em} copy $\x_{n':(n'+J)}$, $t_{n':(n'+J)}$ to cache
			\State \hspace{4em} $\Ddelta_{nn'}:\Ddelta_{nn':(n'+J-1)} \gets (\latentdata_{n} - \latentdata_{n'}):(\latentdata_{n} - \latentdata_{n'+J-1})$ \Comment{vectorized subtraction}
			\State \hspace{4em} calculate $\delta_{nn'}:\delta_{n(n'+J-1)}$  \Comment{vectorized multiplication}
			\State \hspace{4em} calculate $\lambda_{nn'}:\lambda_{n(n'+J-1)}$  \Comment{vectorized evaluation}
			\State \hspace{4em} $\llambda_{n} \gets \llambda_{n} +  \lambda_{nn'}:\lambda_{n(n'+J-1)}$		\Comment{vectorized addition}
			\State \hspace{4em} $n' \gets n' + J$
			\State \hspace{3em} \textbf{end while}
			\State \hspace{2em} \textbf{end for}
			\State \hspace{1em} \textbf{end parfor}
		\end{algsubstates}
		
		\State Compute $N$ $D$-dimensional gradients $\frac{\partial \ell}{\partial \latentdata_n}$:
		\begin{algsubstates}
			\State \hspace{1em} \textbf{parfor} $b \in \{1,\dots,B\}$ \textbf{do}
			\State \hspace{2em} \textbf{if} $b\neq B$ \textbf{then}
			\State \hspace{3em}$Upper \gets b \lfloor  N/B  \rfloor$
			\State \hspace{2em} \textbf{else}
			\State \hspace{3em} $Upper \gets  \lceil  N/B  \rceil$
			\State \hspace{2em} \textbf{end if}
			\State \hspace{2em} \textbf{for} $n \in  \{ (b-1)\lfloor N/B  \rfloor + 1,  \dots,Upper \}$ \textbf{do}
			\State \hspace{3em} copy $\x_{n}$, $t_{n}$ to cache
			\State \hspace{3em} $\frac{\partial \ell}{\partial \latentdata_n} \gets \mathbf{0}$ \Comment{vector of length J}
			\State  \hspace{3em} $n' \gets 1$
			\State  \hspace{3em} \textbf{while} $n' < N$ \textbf{do}
			\State \hspace{4em} $J \gets \min(J,N-n')$
			\State \hspace{4em} copy $\x_{n':(n'+J)}$, $t_{n':(n'+J)}$ to cache
			\State \hspace{4em} $\Ddelta_{nn'}:\Ddelta_{nn':(n'+J-1)} \gets (\latentdata_{n} - \latentdata_{n'}):(\latentdata_{n} - \latentdata_{n'+J-1})$ \Comment{vectorized subtraction}
			\State \hspace{4em} calculate $\delta_{nn'}:\delta_{n(n'+J-1)}$  \Comment{vectorized multiplication}
			\State \hspace{4em} calculate $\mu_{nn'}:\mu_{n(n'+J-1)}$  \Comment{vectorized evaluation}
			\State \hspace{4em} calculate $\xi_{nn'}:\xi_{n(n'+J-1)}$  \Comment{vectorized evaluation}
			\State \hspace{4em} calculate $\xi_{n'n}:\xi_{(n'+J-1)n}$  \Comment{vectorized evaluation}
			\State \hspace{4em} \textbf{for} $j \in {n',\dots,n'+J-1}$ \textbf{do}
			\State \hspace{5em} $\frac{\partial \ell}{\partial \latentdata_n} \gets \frac{\partial \ell}{\partial \latentdata_n}  +  \left( \frac{\mu_{nj}}{\lambda_{n}} + \frac{\mu_{jn}}{\lambda_{j}}\right) \frac{\Ddelta{jn}}{\tau_x^2} +  \left( \frac{\xi_{nj}}{\lambda_{n}} + \frac{\xi_{jn}}{\lambda_{j}}\right) \frac{\Ddelta{jn}}{\sigma_x^2} $
			\State \hspace{4em} \textbf{end for}
			\State \hspace{4em} $n' \gets n' + J$
			\State \hspace{3em} \textbf{end while}
			\State \hspace{2em} \textbf{end for}
			\State \hspace{1em} \textbf{end parfor}
		\end{algsubstates}
	\end{algorithmic}
\end{algorithm}

\begin{algorithm}
	\scriptsize
	\caption{Parallel computation of Hawkes process log-likelihood gradient: 
		\emph{calculates the log-likelihood gradient with multiple levels of parallelization on graphics processing unit (GPU).   In practice, we specify $B=128$ to be the the size of the GPU work groups.  Symbols $\ell$, $\lambda$ and $\Lambda$ appear in Equation \eqref{eq:likelihood}.}}\label{alg:lik}
	\begin{algorithmic}[1]
		\State Compute rates $\lambda_1, \dots, \lambda_N$:
		\begin{algsubstates}
			\State	\hspace{1em}	\textbf{parfor} $n \in \{1,\dots,N\}$ \textbf{do}
			\State\hspace{2em} copy $\x_n$, $t_n$ to local \Comment{$B$ threads}
			\State\hspace{2em}	\textbf{parfor} $N'\in \{1,\dots,\lfloor N/B\rfloor\}$ \textbf{do}
			\State\hspace{3em} $n' \gets N'$
			\State\hspace{3em} $\lambda_{nN'} \gets 0$
			\State\hspace{3em}\textbf{while} $n' < N$ \textbf{do}
			\State\hspace{4em} copy $\x_{n'}$, $t_{n'}$ to local \Comment{$B$ threads}
			\State\hspace{4em} $\Ddelta_{nn'} \gets \latentdata_n - \latentdata_{n'}$ \Comment{vectorized subtraction}
			\State\hspace{4em} calculate $\delta_{nn'} = \sqrt{\sum \Ddelta_{nn'}\circ \Ddelta_{nn'}}$  \Comment{vectorized multiplication}
			\State\hspace{4em} $\lambda_{nN'} \gets \lambda_{nN'}  +\lambda_{nn'}$ \Comment{$\lambda_{nn'}$ a function of $\delta_{nn'}$, $t_n$ and $t_{n'}$}
			\State\hspace{4em} $n' \gets n' + B$
			\State\hspace{3em}	\textbf{end while}
			\State\hspace{2em}\textbf{end parfor}
			\State\hspace{2em} $\lambda_n\gets \sum_{N'} \lambda_{nN'} $   \Comment{binary tree reduction on chip}
			\State\hspace{1em}	\textbf{end parfor}
		\end{algsubstates}
		
		\State Compute $N$ $D$-dimensional gradients $\frac{\partial \ell}{\partial \latentdata_n}$:
		\begin{algsubstates}
			\State	\hspace{1em}	\textbf{parfor} $n \in \{1,\dots,N\}$ \textbf{do}
			\State\hspace{2em} copy $\x_n$, $t_n$ to local \Comment{$B$ threads}
			\State\hspace{2em}	\textbf{parfor} $N'\in \{1,\dots,\lfloor N/B\rfloor\}$ \textbf{do}
			\State\hspace{3em} $n' \gets N'$
			\State\hspace{3em} $\left(\frac{\partial \ell}{\partial \latentdata_n}\right)_{N'} \gets \mathbf{0}$
			\State\hspace{3em}\textbf{while} $n' < N$ \textbf{do}
			\State\hspace{4em} copy $\x_{n'}$, $t_{n'}$ to local \Comment{$B$ threads}
			\State\hspace{4em} $\Ddelta_{nn'} \gets \latentdata_n - \latentdata_{n'}$ \Comment{vectorized subtraction}
			\State\hspace{4em} calculate $\delta_{nn'} = \sqrt{\sum \Ddelta_{nn'}\circ \Ddelta_{nn'}}$  \Comment{vectorized multiplication}
			\State\hspace{4em} $\left(\frac{\partial \ell}{\partial \latentdata_n}\right)_{N'} \gets \left(\frac{\partial \ell}{\partial \latentdata_n}\right)_{N'}  +\left( \frac{\mu_{nn'}}{\lambda_{n}} + \frac{\mu_{n'n}}{\lambda_{j}}\right) \frac{\Ddelta{n'n}}{\tau_x^2} +  \left( \frac{\xi_{nn'}}{\lambda_{n}} + \frac{\xi_{n'n}}{\lambda_{n'}}\right) \frac{\Ddelta{n'n}}{\sigma_x^2} $
			\State\hspace{4em} $n' \gets n' + B$
			\State\hspace{3em}	\textbf{end while}
			\State\hspace{2em}\textbf{end parfor}
			\State\hspace{2em} $\frac{\partial \ell}{\partial \latentdata_n}\gets \sum_{N'}\left(\frac{\partial \ell}{\partial \latentdata_n}\right)_{N'} $   \Comment{binary tree reduction on chip}
			\State\hspace{1em}	\textbf{end parfor}
		\end{algsubstates}
		
	\end{algorithmic}
\end{algorithm}

\section{Additional results}

\begin{figure}[!t]
	\includegraphics[width=\linewidth]{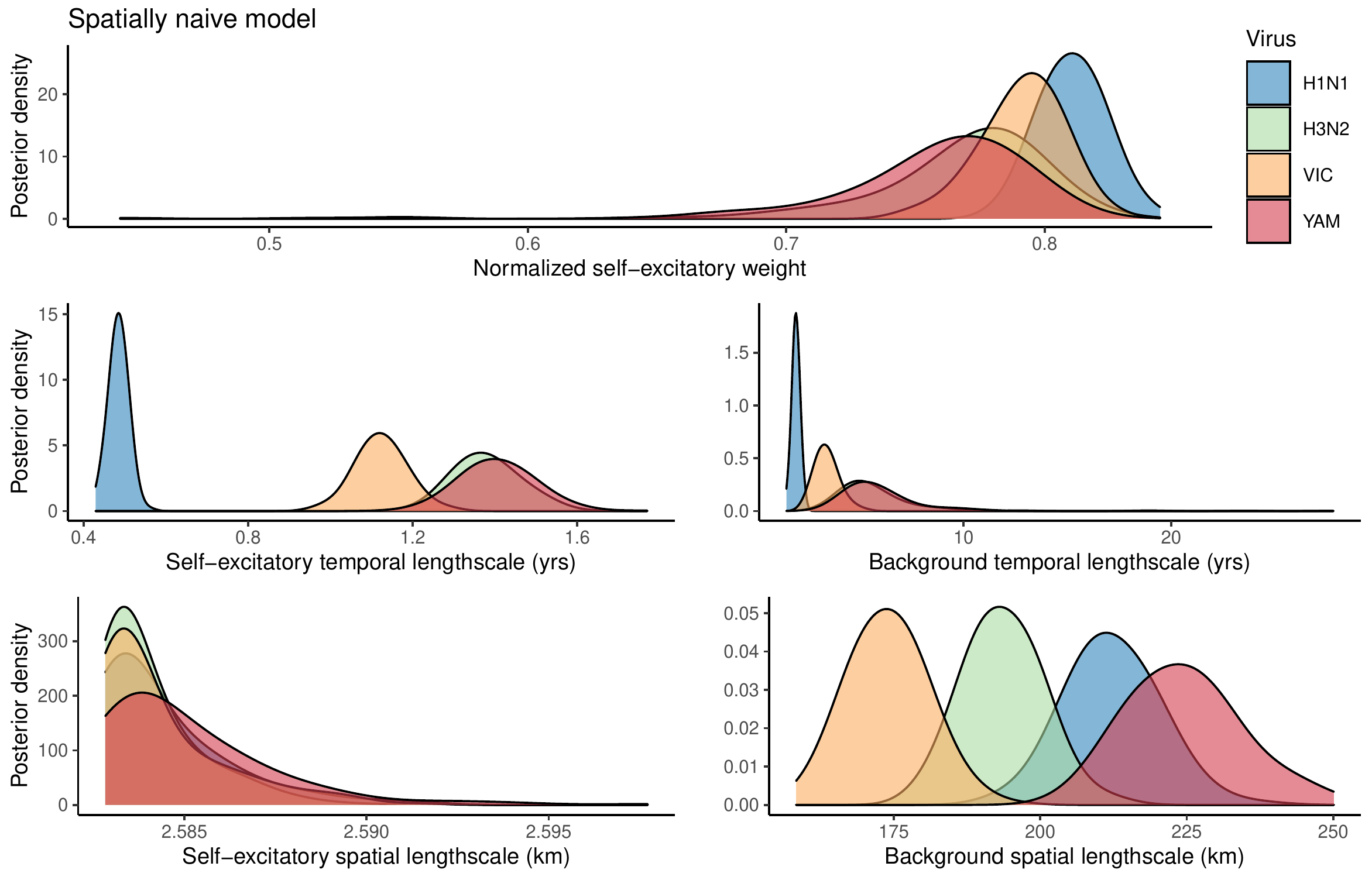}
	\caption{Strain specific posterior inference for a naive model.}\label{fig:naiveInf}
\end{figure}

This section contains posterior inferential results for the naive model from Section \ref{sec:bmdsHawkesDemo} in the form of Figure \ref{fig:naiveInf}.  In this figure, we are particularly interested in the bottom-left plot of posterior densities for self-excitatory spatial lengthscales.  No matter the influenza strain, these posterior distributions concentrate below 3 km.  Such small lengthscales amounts to zero self-excitation between even closely neighboring cities, let alone any transmission on the provincial, national or global scales.  This way the naive model attributes one influenza case (the earliest) to the background process for each distinct location and all successive cases to self-excitation.  One might address this issue by spatially perturbing all cases within, say, the same city, but this would not address the multi-precision nature of the data we consider and certainly would not take any transportation network into account.  In addition, it seems that the resulting spatial lengthscale estimates would depend on the amount of perturbation applied in ways that are difficult to quantify or motivate \emph{a priori}.

\section{Simulation study}\label{sec:sim}

For 800 independent instances, we (a) simulate a spatiotemporal Hawkes process, (b) induce 3 different levels of coarsening and (c) apply our model with both fixed coarsened locations data and inferred locations. We then compare the coverage of resulting credible intervals for both the self-excitatory spatial lengthscale $h$ and the unobserved locations.  Although there is no expectation for Bayesian credible intervals to have perfect frequentist operating characteristics, the model with inferred locations performs reasonably well and outperforms the naive model, especially when coarsening becomes more severe.
	
\begin{figure}
	\centering
	\includegraphics[width=0.6\linewidth]{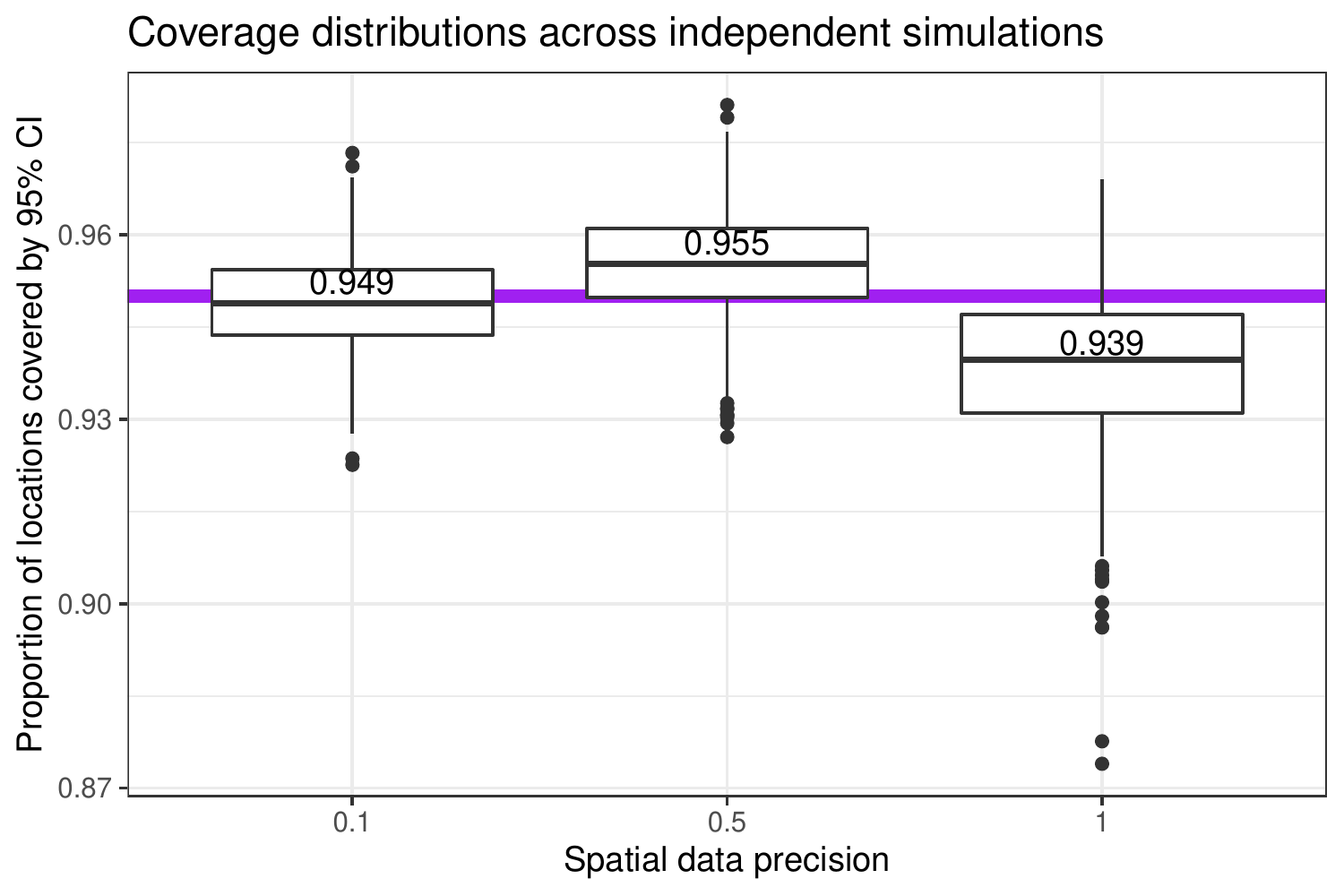}
	\caption{Boxplots and mean proportions of locations covered by 95\% credible intervals (CI) across 800 independent simulations.  The purple line represents 0.95.}\label{fig:sim}
\end{figure}

	To simulate a Hawkes process we use the clustering based algorithm of \citep{zhuang2004analyzing}, which first draws events from an inhomogeneous background process and then simulates successive generations according to the self-excitatory triggering function and its weight.  For the background process, we use a rate function with three Gaussian modes and draw an average of 200 points.  Conditional on these points, we iteratively simulate new generations with an expected number of children of 0.5, and spatial lengthscales of 0.5.  We then coarsen the locations data by rounding to 0.1, 0.5 and 1.0 to investigate performance when the magnitude of coarsening is less than, equal to, and greater than the spatial lengthscale.
	We then generate 30,000 MCMC samples both inferring the latent locations within bounded margins (0.1, 0.5, 1.0) as in Section \ref{sec:dc_uncert} and keeping them fixed. In total, this simulation requires the generation of 30,000 $\times 4 \times 800=$ 96,000,000 MCMC samples, for which we use our Nvidia Quadro GP100 GPU and the algorithms of Section \ref{sec:parallelization}. Table \ref{tab:sim} shows credible interval empirical coverage of the true spatial lengthscale $h$.  In general, we find that coverage is reasonably close to nominal values when inferring locations but deteriorates as a function of the data's spatial precision for the naive model.  Figure \ref{fig:sim} shows the distribution of proportions of locations covered by 95\% credible intervals across the independent simulations.  In general, coverage is reasonably close to 0.95, but deteriorates for data rounded to nearest integer (precision $=1$).  Part of this behavior, including the increased number of outliers, plausibly arises from the need to generate longer MCMC chains.

\begin{table}[!t]
	\centering
	\begin{tabular}{llllllllll}
		\toprule
		& \multicolumn{3}{c}{50\% CIs}& \multicolumn{3}{c}{80\% CIs}& \multicolumn{3}{c}{95\% CIs} \\ \cmidrule(l{2pt}r{2pt}){2-4} \cmidrule(l{2pt}r{2pt}){5-7} \cmidrule(l{2pt}r{2pt}){8-10}
		Spatial precision & 1.0 & 0.5 & 0.1 & 1.0 & 0.5 & 0.1 & 1.0 & 0.5 & 0.1 \\
		\midrule
		Fixed locations & 0.00 & 0.19 & 0.52 & 0.00 & 0.42 & 0.81 & 0.00 & 0.68 & 0.96 \\
		Sampled locations& 0.53 & 0.49 & 0.53 & 0.84 & 0.81 & 0.81 & 0.98 & 0.95 & 0.96 \\
		\bottomrule
	\end{tabular}
	\caption{Empirical coverage from 800 independent simulations for 50\%, 80\% and 95\% credible intervals (CIs) of the self-excitatory spatial lengthscale $h$.}\label{tab:sim}
\end{table}

\end{appendix}

\begin{acks}[Acknowledgments]
 We gratefully acknowledge support from NVIDIA Corporation with the donation of parallel computing resources used for this research.  We also thank Rick Schoenberg and Sudipto Banerjee for their insight and helpful suggestions.
\end{acks}
\begin{funding}
The first author was supported by NIH grant K25 AI153816.  The third author was supported by NIH grant U19 AI135995 and NSF grant DMS1264153.
\end{funding}



\bibliographystyle{imsart-nameyear} 
\bibliography{refs}       


\end{document}